# Understanding the optical properties of doped and undoped 9-armchair graphene nanoribbons in dispersion

*Sebastian Lindenthal[1], Daniele Fazzi[2], Nicolas F. Zorn[1], Abdurrahman Ali El Yumin[1], Simon Settele[1], Britta Weidinger[3], Eva Blasco[3], Jana Zaumseil\*,[1]*

[1]Institute for Physical Chemistry, Heidelberg University, D-69120 Heidelberg, Germany

[2] University of Bologna, Department of Chemistry, 40126, Bologna, Italy

[3] Institute for Molecular Systems Engineering and Advanced Materials and Institute of Organic Chemistry, Heidelberg University, D-69120 Heidelberg, Germany

\*E-mail: zaumseil@uni-heidelberg.de

# ABSTRACT

Graphene nanoribbons are one-dimensional stripes of graphene with width- and edge-structure-dependent electronic properties. They can be synthesized bottom-up in solution to obtain precise ribbon geometries. Here we investigate the optical properties of solution-synthesized 9-armchair graphene nanoribbons (9-aGNRs) that are stabilized as dispersions in organic solvents and further fractioned by liquid cascade centrifugation (LCC). Absorption and photoluminescence spectroscopy reveal two near-infrared absorption and emission peaks whose ratios depend on the LCC fraction. Low-temperature single-nanoribbon photoluminescence spectra suggest the presence of two different nanoribbon species. Based on density functional theory (DFT) and time-dependent DFT calculations, the lowest energy transition can be assigned to pristine 9-aGNRs, while 9-aGNRs with edge-defects, caused by incomplete graphitization, result in more blue-shifted transitions and higher Raman D/G-mode ratios. Hole doping of 9-aGNR dispersions with the electron acceptor $F_4TCNQ$ leads to concentration dependent bleaching and quenching of the main absorption and emission bands and the appearance of redshifted, charge-induced absorption features but no additional emission peaks, thus indicating the formation of polarons instead of the predicted trions (charged excitons) in doped 9-aGNRs.

# INTRODUCTION

Graphene nanoribbons (GNRs) are quasi one-dimensional, nanometre-wide stripes of graphene with an electronic band structure that directly depends on their width and edge geometry.[1-3] Graphene nanoribbons with zigzag edges (zGNRs) show metallic behaviour, while nanoribbons with armchair edges (n-aGNRs) are semiconductors. Here, the n stands for the number of carbon dimer lines perpendicular to the long axis of the GNR and thus indicates the width. Semiconducting aGNRs are interesting as an alternative to graphene and carbon nanotubes for nanoscale optoelectronic devices. Similar to nanotubes, precise control over the structure of the GNRs is required to take advantage of their intrinsic electronic and optical properties.

GNRs can be created with different methods. While top-down techniques, such as electron beam lithography[4] and plasma etching[5] of graphene or unzipping of multi-walled carbon nanotubes,[6] only yield GNRs with a broad width distributions and undefined edge geometries, bottom-up synthesis enables the creation of atomically precise GNRs with a distinct electronic structure. Two different bottom-up approaches are commonly used. The first approach utilizes Ullmann-type coupling of molecular precursors on suitable metal surfaces.[7, 8] This method produces pristine GNRs with high precision but the yield is limited by the substrate surface and subsequent transfer to another substrate is necessary to investigate optical properties or to create nanoelectronic devices.[9, 10] In contrast to that, solution-mediated synthesis can produce large amounts of a wide variety of GNRs with precise width and edge geometry.[11, 12] However, the resulting GNRs are usually polydisperse in length and, due to strong intermolecular interactions, tend to aggregate quickly.[13] Bulky side groups, *e.g.*, branched alkyl chains, along the GNR edges help to prevent aggregation and improve solution processability but can also alter their intrinsic properties.[14-16] Sonication or shear force mixing of GNRs in suitable solvents, *e.g.*, *N*-methylpyrrolidone (NMP), can help to create dispersions with good stability,[17, 18] although film deposition from NMP is difficult due to its high boiling point.

Despite these challenges, the bottom-up synthesis of GNRs has facilitated a number of studies on their electronic and optical properties. The predicted dependence of the bandgap on width was confirmed by scanning tunnelling spectroscopy of different on-surface-synthesized aGNRs.[19] Theoretical and experimental studies revealed that their optical spectra are dominated by excitonic transitions.[20-22] Additionally, photoluminescence (PL) lifetimes on the order of nanoseconds and good photoluminescence quantum yields (PLQY) of up to 80% were reported.[16, 23]

Furthermore, ambipolar charge transport and electroluminescence were observed from bottom-up synthesized aGNRs,[24-26] which raises the question of the impact of charge carriers on the spectroscopic properties of GNRs. Similar low-dimensional materials, such as one-dimensional single-walled carbon nanotubes (SWNTs) or two-dimensional transition metal dichalcogenides (TMDs), exhibit additional optical transitions upon chemical, electrochemical or electrostatic doping.[27-31] Characteristic transitions emerge from the interaction of charges (electrons or holes) with excitonic states in these nanomaterials thus forming quasiparticles – called trions or charged excitons – with red-shifted emission. The trion binding energy (roughly the energy offset between exciton and trion) increases when the dimensionality of the nanomaterial is reduced.[32] Reported values are in the range of a few tens of meV for TMDs[29, 31, 33] and 100-150 meV for SWNTs.[27, 28, 30]

Deilmann *et al.* predicted large trion binding energies on the order of several hundreds of meV for freestanding GNRs.[32] Consequently, the corresponding optical transitions should be observable for doped GNRs. However, the spectroscopic investigation of doped nanoribbons has been hindered by substrate-induced photoluminescence quenching and low yields for on-surface synthesized aGNRs, as well as aggregation issues for solution-synthesized aGNRs. It remains unclear whether solution-synthesized aGNRs can show trionic features upon doping similar to rigid SWNTs or rather exhibit charge-induced quenching and polaron formation due to strong electron-phonon coupling. Polarons are ground state quasiparticles consisting of a charge coupled to a lattice distortion and are typically observed in the absorption spectra of doped polymers. [34, 35]

Here we create stable dispersions of solution-synthesized 9-aGNRs through exfoliation in common and easy to process organic solvents. Further purification by liquid cascade centrifugation (LCC) yields dispersion fractions with slightly different absorption, photoluminescence and Raman spectra indicating the presence of more than one 9-aGNR species. Low-temperature single-nanoribbon spectroscopy and quantum-chemical calculations are used to gain further insights into the structural and optoelectronic properties of these 9-aGNRs and to propose a possible origin of these different species. Finally, the impact of chemical doping with a strong molecular electron acceptor ($F_4$-TCNQ) on the absorption and emission spectra of 9-aGNRs in dispersion is demonstrated and discussed.

## RESULTS AND DISCUSSION

**Synthesis, dispersion and liquid cascade centrifugation of 9-aGNRs**

9-aGNRs (see **Figure 1a**) were selected for this study based on their reported bandgap values ranging from 1.1 to 1.4 eV,[19, 36, 37] resulting in near-infrared (nIR) absorption and emission bands that should still enable the resolution of strongly red-shifted doping-induced features with common InGaAs detectors. The nanoribbons were synthesized *via* a solution-based method previously reported by Li *et al.* (for details see Supporting Information, **Scheme S1**).[36] Briefly, a terphenyl-derivative was polymerized in an AB-type Suzuki reaction to yield the precursor polymer for 9-aGNR (see **Figure 1a**). The length of the polymer, and thus the length of the GNR depends on the number of terphenyl units (marked in red in **Figure 1a**) that are coupled during the polymerization reaction. Size exclusion chromatography (SEC) of the precursor polymer using a polystyrene standard revealed a molecular weight of 30.8 kDa with a polydispersity of 1.46 (see Supporting Information, **Figure S1a**). These values are in good agreement with reports by Li *et al.*[36] and would correspond to an average precursor length of ~25 nm. However, SEC is likely to overestimate the molecular weight of the precursor polymer. Matrix-assisted laser-desorption ionization time-of-flight (MALDI-TOF) mass spectrometry of the precursor polymer only showed peaks up to 17 kDa (see Supporting Information, **Figure S1b**), corresponding to a length of up to 15 nm. Mass spectrometry typically underestimates the molecular weight of polymers, as longer polymer chains are less likely to desorb into the sample chamber and have a higher probability of undergoing fragmentation reactions. Nevertheless, both measurements show that the precursor is polydisperse with a length between 15 and 25 nm. This polydispersity should be retained when the polymer is converted into the 9-aGNR by a Scholl reaction. Atomic force microscopy images of self-assembled 9-aGNR islands on highly ordered pyrolytic graphite (HOPG) show small gaps and reveal that these islands consist of GNRs with different lengths (see Supporting Information, **Figure S2**).

One way to reduce the polydispersity of a nanomaterial dispersion is liquid cascade centrifugation (LCC).[38] In LCC, a stock dispersion is prepared from raw material by sonication in a suitable solvent. It is then subjected to centrifugation steps at increasing rotational centrifugal fields (RCF). Particles that were sedimented during one centrifugation step are collected and redispersed in fresh solvent for the next centrifugation step. Backes *et al.* showed that for 2D materials such as graphene nanosheets or TMDs, LCC leads to a decrease in size and layer number in the sediments for increasing RCF values.[38-40]

Despite previous reports about the instability of dispersions of GNRs without bulky side chains, the as-synthesized 9-aGNR powder could be readily dispersed in tetrahydrofuran (THF) or toluene by simple bath sonication. The resulting stock dispersions were then subjected to centrifugation steps at increasing RCF values of 200, 1000, 10000 and 72000 *g*, yielding five different fractions, which will be referred to by their RCF values (< 0.2 k*g*, 0.2-1 k*g*, 1-10 k*g*, 10-72 k*g*, > 72 k*g*). For a detailed explanation of the LCC process, see **Figure S3** (**Supporting Information**). The 10-72 k*g* fraction was discarded as it barely contained any exfoliated material. Photographs of the other fractions in THF directly after sonication are shown in **Figure 1b**. The colors indicate the amount of exfoliated GNRs in the different fractions, with large quantities in the < 0.2 k*g*, 0.2-1 k*g* and > 72 k*g* fractions. The dispersion stability depends highly on the RCF values during LCC. Differences in dispersion stability and aggregation behaviour become apparent when dispersions are left undisturbed for seven days (see **Figure 1c**). While the < 0.2 k*g* fraction shows almost complete discoloration and a large amount of aggregated GNRs at the bottom of the vial, the appearance of the > 72 k*g* fraction remains unaltered. The latter showed a dispersion stability of up to one year with no changes in appearance or spectroscopic properties.

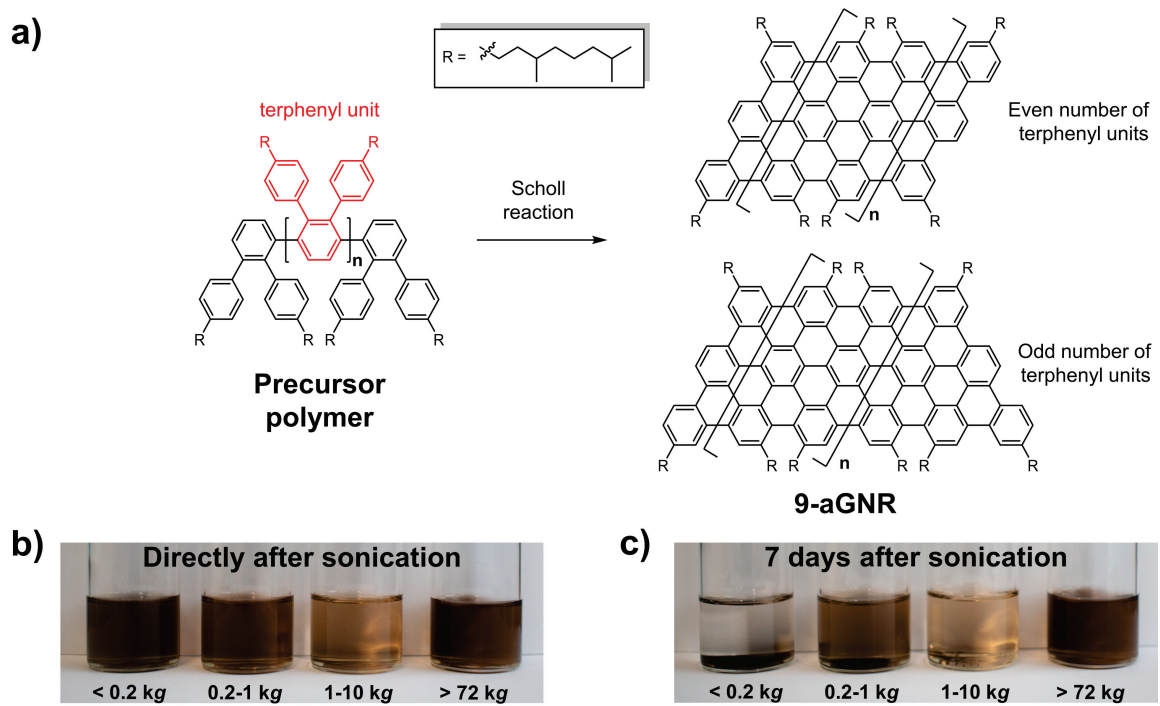

**Figure 1. a)** Bottom-up synthesis of 9-aGNRs *via* a Scholl reaction of the precursor polymer. The precise shape of the 9-aGNRs depends on the number (odd or even) of terphenyl units (red) in the precursor polymer. **b)** Photograph of size-selected dispersions of 9-aGNRs directly after sonication in THF. **c)** Same dispersions of 9-aGNRs left undisturbed for seven days after sonication.

**Spectroscopic characterization of 9-aGNR dispersions**

Due to the challenges of exfoliation and stabilization of solution-synthesized GNRs, little is known about their spectroscopic properties in dispersion. Here, we employed absorption and PL spectroscopy to investigate the influence of LCC on the optical properties of the 9-aGNR dispersions in the neutral state. Baseline-corrected absorption spectra of the different LCC fractions in THF are shown in **Figure 2a**. Strong absorption bands at 350, 477, 526, 817, and 900 nm are apparent that are similar to previously reported absorption spectra of 9-aGNR films[36] but substantially narrower. Absorption spectra of dispersions in toluene showed the same transitions and are presented in **Figure S4** (Supporting Information). The uncorrected spectrum for the <0.2 k$g$ fraction shows a large scattering background (**Figure S5**, Supporting Information), indicating a significant amount of unexfoliated material. A high content of unexfoliated material in the first fraction of LCC is also observed for other nanomaterials, such as TMDs, where it is usually discarded.[40] Absorption spectra measured over 10 hours for the 0.2-1 k$g$ and the >72 k$g$ fractions (**Figure S6**) showed no changes of the observed peaks or the scattering background, thus demonstrating the high dispersion stability of the fractions obtained by LCC.

Interestingly, absorption spectra normalized to the band at 817 nm, show a decreasing contribution of the absorption band at 900 nm for increasing RCF values. The two lowest energy transitions could originate from two different populations of GNRs that exhibit different sedimentation behaviour or they could be intrinsic to a single GNR species and its aggregates. An absorption spectrum with several different transition bands within an energy range of 100-200 meV could be the result of vibronic coupling as reported for polymers or polycyclic aromatic hydrocarbons. However, the ratios of the individual vibronic transitions are determined by the molecular structure and should not change except when aggregates are present. Stronger blue-shifted absorption peaks may indicate the presence of H-aggregates that should be formed by GNRs (see below). LCC usually sorts nanomaterials by their sedimentation coefficients, which are influenced by structural parameters such as size, shape, defectiveness, layer number for TMDs and aggregation tendency.[38, 39] Aggregates should sediment faster and thus should be more prominent in the low RCF fraction, which is the opposite of the observed trend here. Clearly, there is a separation of material with different sedimentation properties between RCF fractions but it is not yet clear, what these differences might be.

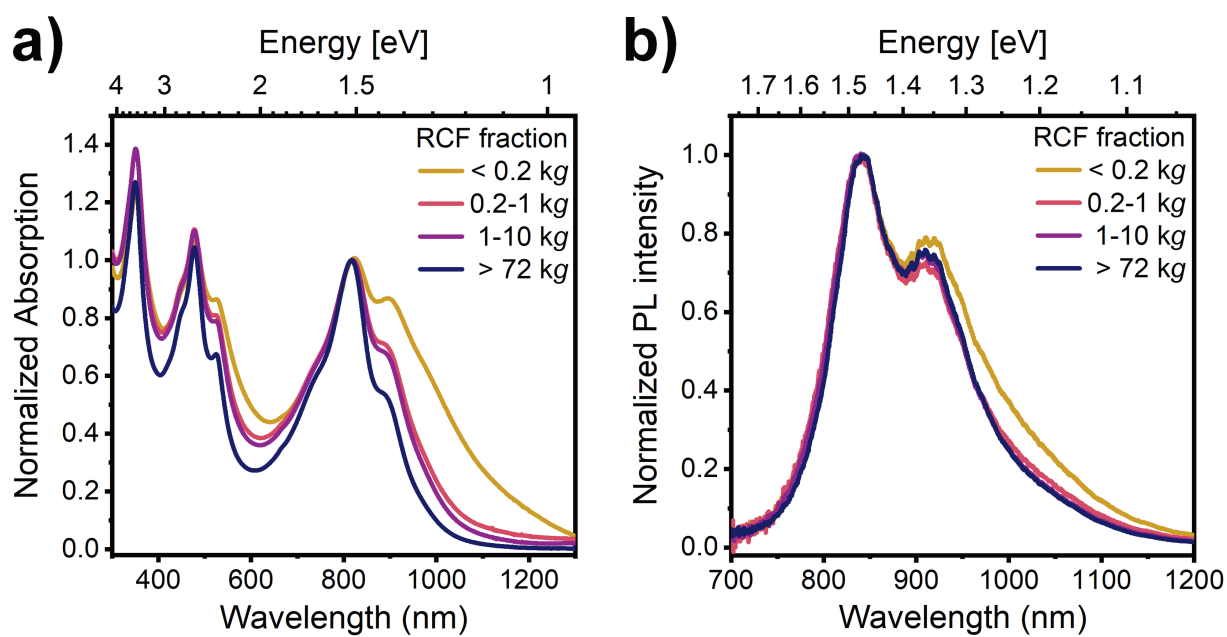

**Figure 2. a)** Absorption spectra of size-selected 9-aGNR dispersions in THF normalized to the absorption band at 817 nm. **b)** PL spectra of size selected 9-aGNR dispersions in THF excited with a pulsed laser at 535 nm and normalized to the emission peak at 840 nm.

Normalized PL spectra of 9-aGNR dispersions in THF excited at 535 nm (see **Figure 2b**) exhibit two emission features at 840 nm and 910 nm. PL spectra of dispersions in toluene are shown in **Figure S7 (Supporting Information)**. The corresponding photoluminescence excitation-emission (PLE) map (**Figure S8**) also shows two emission features at 840 nm and 910 nm that reach their maximum intensity when excited at 535 nm, corresponding well to the observed absorption band at 526 nm. Note that the general shape of the PL spectrum resembles the most redshifted absorption features and is not a mirror image, as would be expected for a vibronic progression. The Stokes shifts of 40 meV for the emission peak at 840 nm and 15 meV for the peak at 910 nm indicate intrinsic PL originating from radiative exciton relaxation (*i.e.*, no excimer emission such as from aggregates). This observation is further corroborated by the fact that time-correlated single photon counting (TCSPC) traces can be fitted with a simple monoexponential decay to determine the PL lifetime (**Figure S9**, Supporting Information). The two different emission features exhibit very similar PL lifetimes between 1.0 ns and 1.2 ns (measured at 840 nm and 920 nm, see **Figure S9**, Supporting Information) that vary slightly but not significantly with RCF values and solvent. The RCF values also do not influence the intensity ratios of the two emission bands, in contrast to the absorption bands. However, there is a clear trend of increasing PLQY for increasing RCF fractions (**Figure S10**). The highest PLQYs are reached for the >72 k*g* fractions with 39 % for dispersions in THF and 71 % for dispersions in toluene. These PLQY values are quite good for organic near-infrared emitters in a wavelength range above 800 nm.[41] The increasing PLQY may also explain the similarities between the PL spectra of the different LCC fractions in **Figure 2b** compared to the differences in the absorption spectra. The PL spectra are likely dominated by the more emissive species and their distribution.

**Low-temperature single-GNR spectroscopy**

A technique that can help to determine whether the two emission features at 840 nm and 910 nm originate from two different GNR populations or both peaks are intrinsic to a single GNR species is low-temperature single molecule spectroscopy. For the case of two different GNR populations with each contributing to only one emission feature, one would expect single-GNR spectra with only one emission peak in the corresponding spectral region. Alternatively, if the two main emission peaks are intrinsic to the 9-aGNR they should always appear together. Measurements at cryogenic temperatures simplify the differentiation between single nanoribbons and aggregates by decreasing linewidths and increasing PL intensities of single fluorophores.[42]

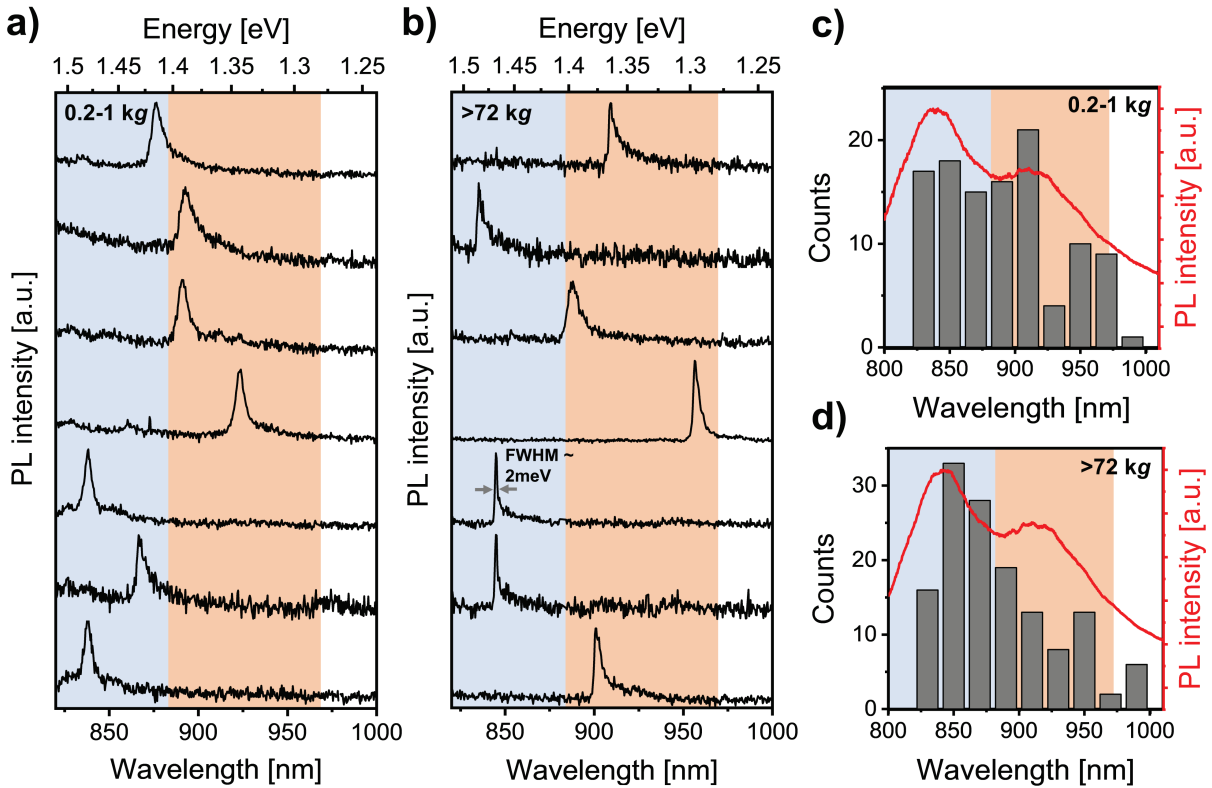

**Figure 3.** PL spectra at 4.6 K of 7 individual 9-aGNRs from the **(a)** 0.2-1 k*g* and **(b)** > 72 k*g* fractions embedded in a polystyrene matrix (excitation wavelength 532 nm). Blue and orange areas highlight the spectral regions associated with the broader emission peaks at 840 nm and 910 nm of the corresponding dispersion at room temperature. **c)** Histogram of peak positions extracted from 108 spectra for the 0.2-1 k*g* fraction. **d)** Histogram of peak positions extracted from 133 spectra for the >72 k*g* fraction. The red lines represent the ensemble PL spectrum of the corresponding 0.2-1 kg and >72 k*g* 9-aGNR dispersions.

Background-corrected PL spectra collected at 4.6 K from several individual 9-aGNRs embedded in a polystyrene matrix (for details see **Experimental Methods**) are shown in **Figure 3a** for the 0.2-1 kg fraction and in **Figure 3b** for the >72 kg fraction. The highlighted areas indicate the spectral regions associated with the peaks at 840 nm (blue) and 920 nm (orange).

For low excitation densities, only one peak per spectrum is observed independent of LCC fraction. The observed emission features are very narrow with the narrowest peaks exhibiting a full-width-at-half-maximum of 2 meV (spectral resolution limit of the setup). Emission spectra measured on the same sample but at room temperature are significantly broader but still show single peaks with substantial variations in position (see **Figure S11, Supporting Information**). The occurrence of only one emission peak for a single nanoribbon indicates that there are indeed different 9-aGNR species that correspond to the two absorption (817 nm and 900 nm) and emission features (840 nm and 910 nm) observed for dispersions.

To ensure that the observed PL signals indeed originated from 9-aGNRs, we recorded over a hundred low-temperature PL spectra (for details see **Experimental Methods**) and determined the positions of the peak maxima. The occurrence of peak maxima at certain wavelengths is plotted in **Figure 3c** and **Figure 3d**. Both histograms show that the majority of peaks is observed in the range between 820 and 940 nm, thus roughly resembling the features of the ensemble PL spectra.

**Raman spectroscopy on 9-aGNR dispersions**

Raman spectroscopy can help to corroborate that all the fractions obtained by LCC contain structurally intact 9-aGNRs. It might also give further insights into the nature of the two GNR populations that are implied by the single-GNR measurements (see above). In addition to the D- and G-modes that are intrinsic to sp²-carbon lattices, graphene nanoribbons exhibit a radial breathing-like mode (RBLM), that decreases in frequency with increasing ribbon width.[43, 44] A width-dependent shear-like mode (SLM) has also been reported for on-surface synthesized 9-aGNRs.[45, 46] **Figure 4a** shows a background-corrected Raman spectrum (excitation laser 785 nm, see **Figure S12a** for spectra without background correction) of a drop-cast film of a 9-aGNR dispersion (for details see **Experimental Methods**). The inset schematically shows the atomic displacement for the RBLM. While wavenumbers between 310 and 316 cm$^{-1}$ have been reported for RBLMs of 9-aGNRs,[10, 37, 45] we observe RBLM at ~280 cm$^{-1}$ (red arrow). This discrepancy might be explained by the presence of the alkyl side chains that were shown to move in phase with the nanoribbon atoms, thus increasing the effective width of the GNR and decreasing the RBLM frequency.[47, 48] Since the RBLM is observed independent of RCF (**Figure S12b, Supporting Information**), all fractions resulting from LCC contain 9-aGNRs. Additionally, a SLM at 188 cm$^{-1}$ is observed for all fractions (blue arrow), which is in good agreement with previously reported experimental (179 cm$^{-1}$) and theoretical (206 cm$^{-1}$) values for 9-aGNRs.[45, 46]

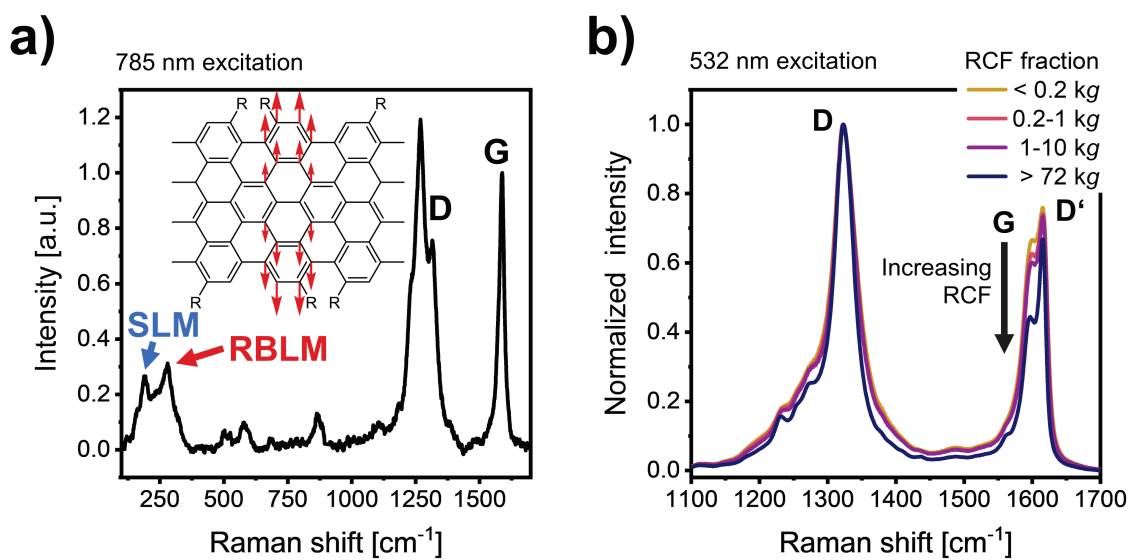

**Figure 4. a)** Baseline-corrected Raman spectrum of a drop-cast film of a >0.2-1 k$g$ 9-aGNR dispersion in THF on glass (excitation 785 nm). The inset shows the atomic displacement for the RBLM of 9-aGNRs. The alkyl chains are represented by -R. **b)** Raman spectra of drop-cast films of different LCC fractions of 9-aGNRs normalized to the D-peak (excitation 532 nm).

Raman spectra of different drop-cast LCC fractions of 9-aGNRs in the wavenumber region above 1000 cm$^{-1}$ show three major peaks at 1322, 1597 and 1615 cm$^{-1}$ (**Figure 4b**) that can be assigned to the D-, G- and D'-mode (for extended Raman spectra up to 3000 cm$^{-1}$ including second order peaks see **Figure S13**, Supporting Information). When normalized to the D-peak, a decrease in G-peak intensity for increasing RCF can be observed. Since the G-mode is intrinsic to the sp²-carbon lattice and the D-mode is only activated near defects, the D/G ratio is often used as a measure for defect density or disorder. High D/G ratios indicate a structurally defective carbon lattice.[49] For aGNRs this correlation is complicated by the fact that armchair edges also activate the D-mode.[49] A high D/G ratio (as in the > 72 kg fraction) could be the result of a larger edge-to-basal plane ratio for shorter GNRs or indeed indicate more defects within the GNR.

Both parameters, GNR length and number of structural defects, may also influence the sedimentation behaviour of the nanoribbons. The attractive forces between graphitic nanomaterials increase with their size, for example, the length for SWNTs or the surface area for graphene.[50] Thus, longer GNRs should have an increased aggregation tendency and sediment faster in LCC. Defects in GNRs that cause structural distortion may prevent aggregation similar to defects in graphene flakes.[51] Assuming similar length distributions, GNRs with structural defects should thus exhibit a higher dispersion stability and slower sedimentation than defect-free GNRs, *i.e.*, they should be more prevalent in higher RCF fractions as indeed observed in the Raman spectra of the different 9-aGNR fractions.

**Quantum chemical calculations regarding shape, size and defects of 9-aGNRs**

The absorption and photoluminescence spectra of the different LCC fractions of 9-aGNRs indicate different populations. Further, the Raman data suggest that these GNR species may exhibit different lengths, shapes or defect densities. To gain deeper insights into the structural and optical properties of 9-aGNRs at the molecular level and to investigate the impact of length, shape and defects, we performed a range of quantum chemical calculations including tight-binding density functional (DFT) semi-empirical methods (GFN2-xTB), DFT and time-dependent DFT (TD-DFT) calculations (for details see **Computational Methods**). In the following, we will refer to the electronic ground state of the GNRs as $S_0$ and to the n-th excited electronic state of the GNR as $S_n$. Transitions between these states are labelled as $S_0$-$S_n$.

First, an oligomer approach was applied by varying the number of repeat units (*i.e.*, terphenyl units) of the 9-aGNRs to examine the dependence of the excited state energies on the length of the nanoribbon and thus to identify the structural model that represents the real system best. For each oligomer, different 9-aGNR shapes were considered, as well as the presence of various types of structural defects. **Figure 5a** shows the optimized structures (GFN2-xTB data, see **Computational Methods**) of the longest oligomers considered in this computational study for two different shapes of 9-aGNRs, *i.e.*, trapezoidal (15 terphenyl units) and parallel (16 terphenyl units.) Trapezoidal and parallel shapes are created by coupling an odd or even number of terphenyl units, respectively. In both cases, the 9-aGNRs are characterized by a flat core structure.

Excited state (singlet) vertical excitation energies were computed at the TD-DFT level for each oligomer length. The strongest dipole-allowed electronic transition, *i.e.*, the one with the highest oscillator strength (*f*), was the $S_0$-$S_1$ transition for each oligomer and for each shape of the nanoribbons, also indicating that the first singlet excited state ($S_1$) is the brightest. Furthermore, there are no dark states (as described at the single-reference TD-DFT level) at a lower energy than $S_1$. The $S_0$-$S_1$ transition can be characterized in terms of the one-particle approximation as the HOMO-LUMO excitation (see **Figure S14**, Supporting Information).

By increasing the oligomer length, the computed $S_0$-$S_1$ transition shifts to lower energies, decreasing from 2.19 eV (short oligomer) to 2.06 eV (longest oligomer, ~ 600 nm, *f* = 7.8) for the parallel shape 9-aGNR, and from 2.23 eV to 2.07 eV (~596 nm, $f$ = 7.13) for the trapezoidal species (see **Figure 5b**). Both types of 9-aGNRs show very similar $S_0$-$S_1$ vertical transition energies, regardless of their shape. By fitting the $S_0$-$S_1$ transition energies of the entire 9-aGNR oligomer series with a model function as previously proposed by Gierschner *et al.*[52] and Kowalczyk *et al.*[53] we obtain an extrapolated *infinite-chain* value of about 1.97 eV (629 nm). This value is very close to the $S_0$-$S_1$ transition energies computed for the longest oligomers considered here (~ 600 nm), thus suggesting that the structural models already reflect electronic properties (*e.g.*, electron delocalization) quite close to the asymptotic (large size) limit. Hence, they can be considered as good representations for the real 9-aGNRs.

Note that the computed $S_0$-$S_1$ transition energies as well as the extrapolated data refer to unscaled TD-DFT values. It is well-known that depending on the choice of the exchange correlation functional and basis set adopted for the calculations (here ωB97X-D/6-31G, see **Computational Methods**), the TD-DFT transition energies are overestimated compared to the experimental values (usually by 0.2-0.6 eV).[54] By rescaling the extrapolated (*infinite-chain*) value (1.97 eV) with respect to the energy of the lowest energy absorption band at 1.37 eV (900 nm, **see Figure 2a**), which is presumably related to the longest 9-aGNR within the sample, we obtain a scaling factor of about 0.70 for the electronic transition energies.

As suggested by quantum-chemical calculations, the shape of the nanoribbon (*i.e.*, parallel vs. trapezoidal) does not affect the $S_0$-$S_1$ optical gap significantly and thus cannot explain the two observed absorption bands (817 and 900 nm, **Figure 2a**). Furthermore, if a substantial length distribution of 9-aGNRs were present in the sample, a spectral broadening should be observed. A very narrow and well-separated bimodal length distribution would be necessary to observe two well-defined peaks instead of a broad absorption and emission feature. Such a bimodal distribution is, however, not apparent in SEC or mass spectra of the precursor polymer (see

**Figure S1**), indicating that the two different transitions are most likely not the result of GNR populations with different lengths. Hence, other possible factors must be considered.

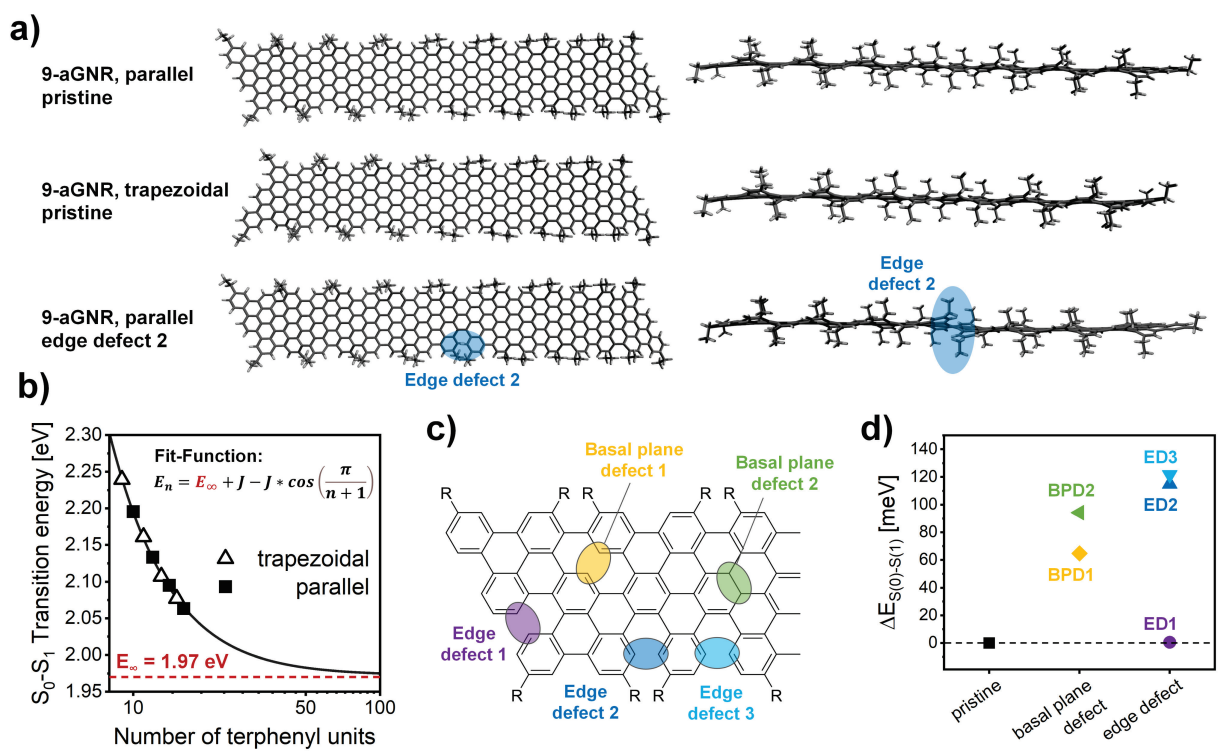

**Figure 5. a)** GFN2-xTB optimized structures for pristine (top and middle) 9-aGNRs with 16 and 15 terphenyl units, respectively, and a defected (ED2) 9-aGNR (bottom). **b)** Unscaled TD-DFT energies for the lowest bright transition ($S_0$-$S_1$) versus oligomer length (parallel shape: filled squares, trapezoidal shape: open triangles) and corresponding fit. **c)** Schematic depiction of structural defects that can occur during a Scholl reaction. **d)** Energy difference (TD-DFT data) between the $S_0$-$S_1$ transition energy computed for defected 9-aGNR species and the pristine case (parallel shape, 16 terphenyl units). See Supporting Information for the corresponding values for trapezoidal shape.

Several studies have attributed the emergence of two distinct peaks in absorption and PL spectra to GNR aggregation and inter-GNR energy transfer.[16, 55, 56] Thus, we also calculated the ground to excited state (singlet) transition energies for a 9-aGNR monomer, consisting of 10 terphenyl units, and its van-der-Waals dimer. Given the large size of the dimer (> 600 atoms), a full geometry optimization was possible only at the GFN2-xTB level of theory, while the excited state energies were computed at the semi-empirical sTD-DFT level. According to these dimer-based calculation, 9-aGNRs should form H-type aggregates resulting in blue-shifted transition energies (see **Figure S15**, Supporting Information). Since aggregates sediment at lower RCF values than monomers,[39, 40] larger amounts of H-aggregates should be found in low RCF fractions and lead to an increase in optical density of the peak at 817 nm. However, **Figure 2a** shows the opposite trend. The relative contribution of the absorption band at 817 nm is lower for lower RCF values, thus rendering aggregates improbable as the origin for the peak at 817 nm.

Another possible cause for different transition energies might be defects within the 9-aGNR core. The graphitization of the precursor polymer to the 9-aGNR was achieved by a Scholl reaction with 2,3-dichloro-5,6-dicyano-3,4-benzoquinone (DDQ) as an oxidative agent. Scholl reactions can be highly efficient with yields near 100 % under ideal conditions.[57] However, small amounts of incompletely graphitized GNRs are possible and will remain undetected by the usual analytical techniques. Based on the estimated length of the precursor polymer the final 9-aGNR should contain at least ~25 terphenyl units. Four bonds have to be closed for each terphenyl unit during the Scholl reaction. Thus, in a GNR with 25 terphenyl units, 100 additional bonds have to be formed. With an efficiency of 99% as determined by Li *et al.* for this reaction,[36] a typical GNR may still contain one unclosed bond, which would constitute a defect in the $sp^2$-hybridized lattice.

Consequently, we investigated the presence of different structural defects in our model-systems and their effect on the electronic transition energies. For the longest 9-aGNR oligomers (parallel and trapezoidal shapes) we considered five defect types. The molecular structures that could result from missing bonds at different positions and hence different defect types are shown in

**Figure 5c**. Note that edge defects (ED) are more probable than basal plane defects (BPD) as Scholl reactions are more efficient for pre-oriented precursors.[57] The optimized molecular structures (GFN2-xTB level) for each defected 9-aGNR species are reported in the Supporting Information (**Figure S16**). Generally, the introduction of an edge defect has a more substantial impact on the structure of the nanoribbon as it induces a significant twist to the backbone. This twist is particularly evident for ED2 and ED3, which are located on the longitudinal edge of the nanoribbon and thus perturb the effective conjugation along the GNR core. The presence of a basal plane defect (BPD1 and BPD2) on the other hand is less detrimental and does not induce large structural distortions.

The structural reorganizations introduced by each defect affect the electron delocalization and thus also the $S_0$-$S_1$ optical gap. For each defect-type and for both parallel and trapezoidal 9-aGNR species, we computed the excited state energies at the TD-DFT level (see **Figure 5d** for parallel and **Figure S17** for trapezoidal GNRs). As already implied by the structural deformations, the defected 9-aGNRs show larger $S_0$-$S_1$ transition energies than the pristine species. Edge defects, in particular ED2 and ED3, induce an increase of the optical gap of circa 0.11-0.13 eV with respect to the pristine ribbons. ED1, being located on the terminal edge, does not perturb the electron conjugation of the plane as much as ED2 and ED3 and the optical gap remains similar to the pristine species. BPD1 and BPD2 defects also induce a blue-shift of the $S_0$-$S_1$ transition, which is however lower in magnitude (about 0.06 - 0.1 eV) than for ED2 and ED3. The $S_0$-$S_1$ transitions exhibit similar oscillator strengths for pristine and defected 9aGNRs (**Figure S17** and **S18**, **Supporting Information**).

Our calculations show that the presence of defects changes the electronic transition energies, modulating the optical gap and possibly causing the shift of the absorption and emission bands of 9-aGNRs. These results imply that the strong transition observed at 817 nm (1.5 eV, **Figure 2a**) may originate mostly from defective 9-aGNRs with presumably ED2 or ED3 type defects, while the transition at 900 nm (1.37 eV) could be assigned to defect-free or ED1-containing 9-aGNRs. The experimentally observed energy difference between the two absorption bands is 0.13 eV, which is close to the computed $S_0$-$S_1$ energy offset (**Figure 5d**).

Similar to defects in graphene[49] and SWNTs,[58] ED2 and ED3 defects could also act as additional electron-defect scattering sites. They should lead to a higher efficiency of one-phonon defect-assisted processes and an increase in relative intensities of the D- and D'-bands in Raman spectra. LCC fractions with higher D/G-ratios also exhibit a higher relative contribution of the absorption peak at 820 nm (**Figure 4b**), which is consistent with the proposed assignment. Note that the G- and D'-bands are nicely separated even in the >72 kg fraction indicating that a large fraction of the graphene lattice is still defect-free. Thus, in contrast to SWNTs and graphene, the steady increase in D/G ratios for increasing RCF values should not be interpreted as a larger number of defects per GNR but rather as an increasing number of GNRs that contain ED2 or ED3 defects.

In summary, the combination of experimental data and quantum chemical calculations clearly points toward defective nanoribbons (most likely containing edge defects) as one major population within our dispersions in addition to pristine 9-aGNRs. These different GNR species can explain the dependence of the two absorption and emission peaks as well as the Raman D/G ratios on LCC fractions (*i.e.*, different RCF values).

**Chemical doping of 9-aGNR dispersions with $F_4TCNQ$**

After gaining a better understanding of the optical properties of undoped 9-aGNRs in dispersion, we now investigate the impact of excess charge carriers on these properties. Chemically doped SWNTs show relatively narrow charge-induced absorption and emission features typically associated with trions.[27] However, when conjugated polymers are chemically or electrochemically doped, very broad red-shifted polaron transitions appear in their absorption spectra, reflecting changes in the electronic structure of the polymer upon charging.[34, 35, 59, 60]

2,3,5,6-Tetrafluoro-7,7,8,8-tetracyanoquinodimethane (F$_4$TCNQ) is a strong molecular electron acceptor that has been used previously to induce p-doping in polymers,[61] semiconducting SWNTs,[62] and graphene[63] by forming charge-transfer complexes and was chosen here for direct doping of nanoribbons in dispersion. As F$_4$TCNQ quickly decomposes in THF, only 9-aGNR dispersions in toluene could be doped with this method (for details, see **Experimental Methods**). **Figure 6a** shows absorption spectra of a >72 k$g$ fraction with increasing concentrations of F$_4$TCNQ. For low F$_4$TCNQ concentrations (2 and 5 µg mL$^{-1}$) the absorption band at 900 nm starts to be bleached and an additional redshifted charge-induced absorption feature at 1300 nm emerges (indicated by the arrow in **Figure 6a**). While this feature is fairly weak for the >72 k$g$ fraction, it appears stronger for doping of the 0.2-1 k$g$ fraction (see **Figure S19a, Supporting Information**). Since the undoped 0.2-1 k$g$ fraction in toluene also shows a stronger absorption band at 900 nm than the >72 k$g$ fraction, it is reasonable to assume that the additional redshifted absorption feature at 1300 nm is correlated with this transition. For higher dopant concentrations, bleaching of the absorption band at 817 nm commences and another charge-induced redshifted absorption feature at 1100 nm appears. The presence of an isosbestic point (at ~900 nm) clearly shows that the two features are connected. Interestingly, for the highest doping concentrations, the feature at 1300 nm bleaches again. For the 0.2-1 k$g$ fraction, we observe an even further redshifted absorption at 1600 nm and above (**see Figure S20a, Supporting Information**). Note that the absorption band at 900 nm shows stronger bleaching than that at 817 nm for the same F$_4$TCNQ concentrations, indicating again that there must be two different 9-aGNR species with different redox potentials.

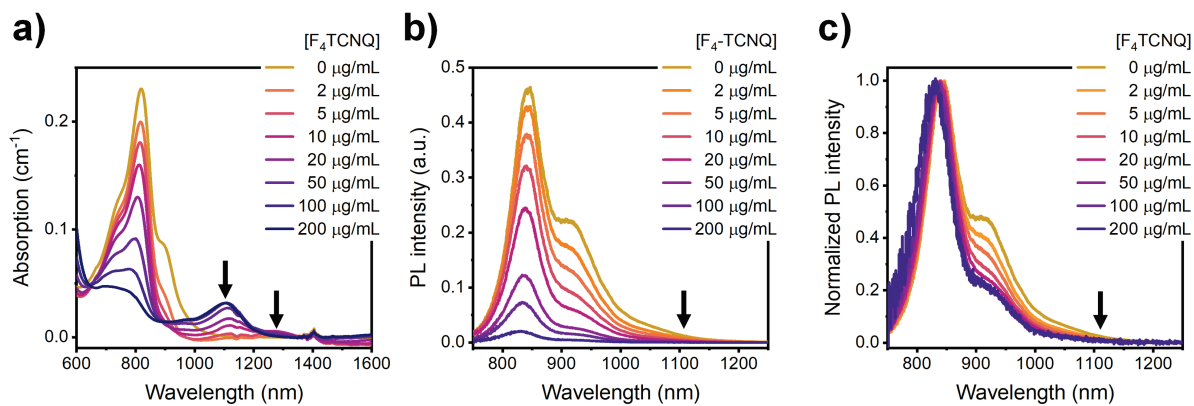

**Figure 6. a)** Absorption spectra of a >72 k$g$ 9-aGNR dispersion doped with different concentrations of F$_4$TCNQ. Positions of charge-induced absorption peaks are marked by arrows. **b)** PL spectra (absolute values) of a 9-aGNR dispersion doped with F$_4$TCNQ and **c)** normalized PL spectra. Arrows indicate the position where trion emission would be expected according to the charge-induced features in the absorption spectra.

To differentiate between trions and polarons as the origin of the redshifted, charge-induced absorption, PL spectra of the F$_4$TCNQ-doped 9-aGNR dispersions were recorded. As the trion is an emissive excited state (charged exciton), we would expect to observe redshifted emission peaks at 1100 nm and 1300 nm. **Figure 6b** shows the PL spectra of a >72 k$g$ fraction doped with increasing amounts of F$_4$TCNQ. The arrow indicates the position of the observed charge-induced absorption band. However, for increasing dopant concentrations, we only observe the expected quenching of the main emission and no additional emission features. Spectra showing the region up to 1400 nm are depicted in **Figures S19b** and **S20** (Supporting Information) and also do not show any further red-shifted emission. In PL spectra normalized to the emission peak at 840 nm (**Figure 6c**), a blue-shift with increasing doping levels is apparent, which is reminiscent of the blue-shift observed for SWNTs upon doping.[64] In addition, we find that the emission at 910 nm is quenched more strongly than that at 840 nm for the same dopant concentration, again indicating the presence of two different species with different redox potentials.

These findings strongly suggest polaron formation upon doping of solution-synthesized 9-aGNRs in dispersion and seem to exclude the possibility of trion formation. This is in stark contrast to the recently reported trionic PL by Fedotov *et al.* of on-surface synthesized 7-aGNRs.[65] A possible explanation could be the stronger deformation of solution-synthesized 9-aGNRs due to the sterically demanding branched alkyl side chains that may promote charge localization and stronger electron-phonon coupling.

To further understand the observed spectra of p-doped nanoribbons, quantum chemical calculations were also performed for charged 9-aGNRs. The singly charged states (radical cation, +1) of both parallel and trapezoidal 9-aGNRs were optimized at the GFN2-xTB level. The electronic transitions were further computed at the TD-DFT level adopting a spin-polarized unrestricted (UωB97XD) functional. Furthermore, a comparison between pristine and defected 9-aGNRs is reported in the Supporting Information (**Figure S21**). For each species, electronic transitions at lower energies (red-shifted) than the $S_0$-$S_1$ optical gap of the undoped nanoribbon appeared. The charge-induced (dipole allowed) electronic transitions are computed at about 0.5-0.8 eV below the $S_0$-$S_1$ transition of the undoped species. This energy offset is comparable (within the approximations of GFN2-xTB and TD-DFT methods) to the experimental values (0.38 and 0.42 eV). The presence of defects or different nanoribbon shapes slightly affect the transition energies of the charged species (**Figure S21**), suggesting a localization of the polaron spin density on the plane of the nanoribbon. As shown above, the nanoribbon shape and defects influence the conjugation length, thus also changing the delocalization of the polaron spin density.

Note that the presence of defects also changes the energy of the frontier molecular orbitals (*e.g.*, HOMO and LUMO levels), thus affecting in a first approximation (*i.e.*, following Koopman's theorem, *IP(EA) = -$\varepsilon_{HOMO}$(-$\varepsilon_{LUMO}$)*) the redox potentials as well. For example, the DFT-computed HOMO energy for 9-aGNRs with ED2 and ED3 defects (-5.6 eV) is lower than that of the pristine species (-5.5 eV). BPD1 and BPD2 behave similarly to EDs, however with a reduced energy shift of only 0.08 eV instead of 0.1 eV. These calculations again can help to understand the experimental doping data. Efficient p-doping takes place if the energy difference $\Delta E$ between the HOMO of the GNR and the LUMO of the dopant (here $F_4TCNQ$) is large. Thus, pristine 9-aGNRs (assumed to correspond to the absorption band at 900 nm) with a higher-lying HOMO level (larger $\Delta E$) should be oxidized (p-doped), and thus bleached at lower $F_4TCNQ$ concentrations than the defective 9-aGNRs (which are assigned to the band at 817 nm). The defective 9-aGNRs are oxidized at higher $F_4TCNQ$ concentrations due to their lower-lying HOMO (smaller $\Delta E$) compared to the LUMO of $F_4TCNQ$. Thus, the computed HOMO energies of defective and pristine 9-aGNRs further corroborate the corresponding assignment of the two absorption and emission peaks as described above.

## CONCLUSION

In this study we created stable dispersions of solution-synthesized 9-aGNRs in organic solvents and revealed the origin of their absorption and emission features in the neutral and hole-doped state. We found that the as-synthesized raw material contains two species of 9-aGNRs that result in two different absorption and emission bands in the near-infrared. Based on semiempirical and DFT/TD-DFT calculations we could assign the lowest energy band to 9-aGNRs without defects and the dominant blue-shifted band to 9-aGNRs with edge-defects. The ratio of the two nanoribbon populations varies between different LCC fractions. Molecular p-doping of these 9-aGNR dispersions with $F_4TCNQ$ resulted in fairly narrow redshifted charge-induced absorption features and PL quenching but no charge-induced emission that would indicate the existence of trions. We conclude, that solution-synthesized 9-aGNRs in dispersion favour polaron formation similar to conjugated polymers. In contrast to theoretical predictions, the flexibility and torsion of GNRs with alkyl sidechains in dispersion seem to prevent the formation of stable and emissive trion states.

## EXPERIMENTAL METHODS

**Synthesis of 9-aGNR.** Atomically precise 9-aGNRs were synthesized according to an adapted protocol by Li *et al.* [36] and described in detail in the supporting information (see **Figure S1** for reaction scheme).

**Preparation of 9-aGNR Dispersions.** 10 mg of 9-aGNR powder were added to 10 mL of solvent (THF or toluene) in a 25 mL round flask which was then sealed with a septum. This mixture was ultrasonicated for 4 hours in a Branson 2510 sonication bath during which the temperature was held constant at room temperature.

**Sorting of 9-aGNRs by LCC.** A freshly prepared 9-aGNR dispersion was exposed to increasing rotational centrifugal forces (RCF). The samples were centrifuged at 200 $g$, 1000 $g$ and 10000 $g$ in a Hettich Mikro 220R centrifuge, equipped with a 11.95A fixed-angle rotor. The supernatant of the 10000 $g$ centrifugation step was centrifuged at 72000 $g$ using a Beckmann Coulter Avanti J-26S XP centrifuge, equipped with a JA25.50 fixed-angle rotor. This process yields 5 fractions with GNRs of different size (< 200 $g$, 200-1000 $g$, 1000-10000 $g$, 10000-72000 $g$ and >72000 $g$).

**Chemical Doping with $F_4TCNQ$.** A stock solution of $F_4TCNQ$ in toluene with a concentration of 1 mg mL$^{-1}$ was prepared. To achieve the desired doping level, a suitable amount of $F_4TCNQ$ stock solution and pure toluene were added to a size-selected 9-aGNR dispersion in toluene so that in the final mixture the optical density of the absorption peak at 817 nm is at 0.2 cm$^{-1}$.

**Raman spectroscopy.** Raman spectra of drop-cast GNR dispersions were acquired with a Renishaw inVia confocal Raman microscope in backscattering configuration equipped with a 50× long working distance objective (N.A. 0.5, Olympus). To minimize the influence of spot-to-spot variations, maps with >500 spectra were recorded and averaged. For excitation at 785 nm, the obtained Raman spectra were baseline-corrected to account for the PL background.

**Absorption spectroscopy.** Baseline-corrected absorption spectra were recorded using a Cary 6000i UV-Vis-NIR absorption spectrometer (Varian, Inc.) and cuvettes with 1 cm path length.

**Photoluminescence spectroscopy.** PL spectra were acquired from dispersion by excitation at 532 nm (picosecond-pulsed supercontinuum laser (NKT Photonics SuperK Extreme). Emitted photons were collected by a NIR-optimized 50× objective (N.A. 0.65, Olympus) and spectra were recorded with an Acton SpectraPro SP2358 spectrometer with a liquid-nitrogen-cooled InGaAs line camera (Princeton Instruments, OMA-V:1024).

**PL Quantum Yield Measurements.** The PLQY of the sample was determined by an absolute method using an integrating sphere as described in detail elsewhere. [66] The dispersion (with optical density of ~ 0.1 cm$^{-1}$ at 817 nm) in a quartz cuvette was placed inside the integrating sphere and excited at 535 nm by the spectrally-filtered output of a picosecond pulsed supercontinuum laser source. The absorption of laser light and the fluorescence were transmitted to the spectrograph *via* an optical fiber. The measurement was repeated with pure solvent to account for scattering and absorption of the solvent and the cuvette.

**Lifetime Measurements.** Photoluminescence lifetimes were measured by time-correlated single photon counting (TCSPC). The emission from the dispersion (excited at 535 nm) was spectrally filtered by an Acton SpectraPro SP2358 spectrograph and focused onto a gated silicon avalanche photodiode (Micro Photon Devices). Arrival times of photons were recorded with a time-correlated single photon module (Picoharp 300, Picoquant). The fluorescence decay histograms were fitted by a monoexponential fit procedure.

**Low-temperature single GNR spectroscopy.** For low-temperature single GNR spectroscopy, a GNR dispersion in THF was diluted to an optical density of 0.002 at 817 nm and mixed with the same volume of a 40 mg mL$^{-1}$ polystyrene solution in THF. 15 μL of this mixture were then spin-coated (2000 rpm, 1 min) onto a glass slide coated with 150 nm gold. The resulting samples were mounted inside a liquid helium-cooled closed-cycle cryostat (Montana Instruments Cryostation s50) and cooled to 4.6 K under high vacuum conditions ($10^{-5}$ - $10^{-6}$ bar). The output of a continuous wave laser diode (Coherent, Inc. OBIS 532 nm) was focused onto the sample using a 50× long working distance objective (Mitutoyo, N.A. 0.42). Scattered laser light was blocked by using appropriate long-pass filters. Spectra were recorded with a 1340x400 Si CCD Camera (Princeton Instruments, PIXIS:400) coupled to a grating spectrograph (Princeton Instruments IsoPlane SCT 320) using a grating with 150 grooves mm$^{-1}$ and 800 nm blaze. For each spot at least 5 spectra with an integration time of one minute were recorded and averaged. To record individual spectra for the creation of histograms, an area of 99 x 99 μm was scanned with a step width of 3 μm. Emission peaks were observed for 108 (0.2-1 k*g* fraction) or 133 (>72 k*g* fraction) spectra out of a total of 1156 measured spots.

## COMPUTATIONAL METHODS

**Semiempirical (GFNn-xTB) and DFT calculations.** 9-aGNR species were modelled by adopting an oligomer approach and changing the chain length, namely the number of terphenyl units, of the nanoribbons. Oligomers containing eight to sixteen terphenyl units were considered. Different shapes of the 9-aGNRs were investigated, namely parallel and trapezoidal, featuring an even or an odd number of terphenyl units, respectively. For each oligomer and for each shape, the molecular structure, the electronic structure and the optical transitions (*i.e.*, the ground to excited state (singlet) vertical excitations) were calculated at the semi-empirical (*i.e.*, GFN2-xTB, sTD-DFT), and DFT, TD-DFT levels of theory (see below). To reach an effective balance between computational cost and accuracy, the longest 9-aGNRs investigated here comprised 16 (parallel shape) or 15 terphenyl units (trapezoidal shape), respectively. For both of them, various structural defects were further considered, *i.e.*, edge defects (three different types, ED1, ED2 and ED3, see **Figure 5**) and basal defects (two kinds, BPD1 and BPD2, see **Figure 5**). The presence of alkyl side chains was included, however the length of each chain was reduced with respect to the synthesized 9-aGNRs in order to limit the computational costs.

All geometries were fully optimized at the GFN2-xTB level by adopting very tight thresholds for energy and gradients convergence.[67, 68] For some oligomers, extra DFT geometry optimizations were performed, leading to results very similar to the GFN2-xTB structures. Vertical electronic transitions from ground to singlet excited states were computed both with the semi-empirical sTD-DFT approach[67] and TD-DFT. Vibronic effects were not considered in these calculations.

For DFT and TD-DFT calculations the range-separated functional with the inclusion of dispersion corrections ωB97X-D and the Pople double split-valence basis set with diffusion functions (6-31G*) were used.

The molecular and electronic structures of charged species, *i.e.*, radical-cations (+1), were optimized at the GFN2-xTB level, while the vertical electronic excited state energies were computed at the TD-UDFT level adopting the spin polarized unrestricted approach. Charged species were computed for the longest 9-aGNR oligomers (parallel and trapezoidal shape) without (pristine) and with structural defects.

GFN2-xTB calculations were performed with the open-source code xTB (v 6.4.1),[68] while DFT and TD-DFT calculations were carried out with Gaussian 16[69] or ORCA[70] (v.5.0.3) programs.

The model function used to extrapolate the oligomer $S_0$-$S_1$ (TD-DFT) unscaled excitation energies (**Figure 5b**) was proposed by Kowalczyk *et al.*[53] It refers to the Frenkel exciton model and reads as:

$$E_n = E_\infty + J\left(1 - \cos\left(\frac{\pi}{n+1}\right)\right)$$

with $E_\infty$ as the extrapolated value for an infinite chain length (n) and J as the coupling constant.

**Author Contributions**

S.L. synthesized the 9-aGNRs, processed and measured all samples and analyzed the data. D.F. performed and analyzed the quantum-chemical calculations. S.S. and N.F.Z. contributed to the characterization of dispersions and films. A.A.E.Y. and N.F.Z. contributed to low-temperature PL measurements and data analysis. B.W. supervised by E.B. performed size-exclusion chromatography on the precursor polymer. J.Z. conceived and supervised the project. S.L., D.F. and J.Z. wrote the manuscript with input from all authors. All authors have given approval to the final version of the manuscript.

**Notes**

The authors declare no competing financial interest.

ACKNOWLEDGMENT

This project has received funding from the European Research Council (ERC) under the European Union's Horizon 2020 research and innovation programme (Grant Agreement No. 817494 "TRIFECTs"). D. F. acknowledges partial funding from the National Recovery and Resilience Plan (NRRP), Mission 04 Component 2, Investment 1.5 – NextGenerationEU, Call for tender n. 3277 dated 30/12/2021, Award Number: 0001052 dated 23/06/2022. E.B. acknowledges funding from the Deutsche Forschungsgemeinschaft (DFG, German Research Foundation) *via* the Excellence Cluster "3D Matter Made to Order" (EXC-2082/1-390761711 and the Carl Zeiss Foundation through the Carl-Zeiss-Foundation-Focus@HEiKA. S.L. and J.Z. warmly thank Milan Kivala and his team for providing laboratory space and advise on the GNR synthesis.

# SUPPORTING INFORMATION

# Understanding the optical properties of doped and undoped 9-armchair graphene nanoribbons in dispersion

*Sebastian Lindenthal[1], Daniele Fazzi[2], Nicolas F. Zorn[1], Abdurrahman Ali El Yumin[1], Simon Settele[1], Britta Weidinger[3], Eva Blasco[3], Jana Zaumseil*,[1]*

[1]Institute for Physical Chemistry, Heidelberg University, D-69120 Heidelberg, Germany
E-mail: zaumseil@uni-heidelberg.de

[2] University of Bologna, Department of Chemistry, 40126, Bologna, Italy

[3] Institute for Molecular Systems Engineering and Advanced Materials and Institute of Organic Chemistry, Heidelberg University, D-69120 Heidelberg, Germany

*E-mail: zaumseil@uni-heidelberg.de

# Contents

## Synthesis of 9-aGNRs

### General

The synthesis of 9-aGNRs follows the published method by Li *et al.*[1] with some minor modifications.

Reagent grade chemicals were purchased from Sigma Aldrich, Alfa Aesar, Tokio Chemical Industries or Fisher Scientific and used without further purification. Anhydrous solvents were obtained from an MBraun MB SPS-800 solvent purification system, with the exception of dry THF, which was freshly distilled over sodium prior to usage. All reactions containing moisture or air-sensitive components were carried out in inert atmosphere and in dry reaction vessels with standard Schlenk techniques unless noted otherwise.

For analytical thin-layer chromatography (TLC) aluminum plates coated with 0.2 mm of silica gel and fluorescent indicator (Macherey-Nagel, ALUGRAM®, SIL G/UV$_{254}$) were used. Spots on TLC plates were observed by exposure to UV light ($\lambda$ = 254 nm and 366 nm). Column chromatography was performed on silica gel (Macherey-Nagel, M-N Silica Gel 60A, 230-400 mesh).

$^1$H and $^{13}$C NMR spectra were recorded either on a Bruker Avance DRX 300 (300 MHz for $^1$H and 75 MHz for $^{13}$C), Bruker Avance III 400 (400 MHz for $^1$H and 100 MHz for $^{13}$C) or Bruker Avance III 600 (600 MHz for $^1$H and 150 MHz for $^{13}$C) spectrometer. Chemical shifts $\delta$ are reported in ppm and referenced to the residual solvent signal ($\delta_H$(CHCl$_3$) = 7.26 ppm, $\delta_C$(CHCl$_3$) = 77.2 ppm). Coupling constants (J) are given in Hz. To report the multiplicity the following abbreviations were used: s = singlet, d = doublet and m = multiplet for the $^1$H NMR and s = primary, d = secondary, t = tertiary and q = quarternary for $^{13}$C NMR.

GPC measurements were performed on a Shimadzu Nexera LC-40 system (with LC-40D pump, autosampler SIL-40C, DGU-403 (degasser), CBM-40 (controlling unit), column oven CTO-40C, UV-detector SPD40 and RI-detector RID-20A). The system was equipped with 4 analytical GPC-columns (PSS): 1 x SDV precolumn 3 μm 8x50 mm, 2 x SDV column 3 μm 1000Å 8x300 mm, 1 x SDV column 3 μm 10$^4$Å 8x300 mm. The measurements were performed in THF at a flow speed of 1 mL/min at a temperature of 40 °C. Chromatograms were analyzed using the LabSolutions (Shimadzu) software. Calibration was performed against different or polystyrene standards (370 - 2 520 000 Da, PSS).

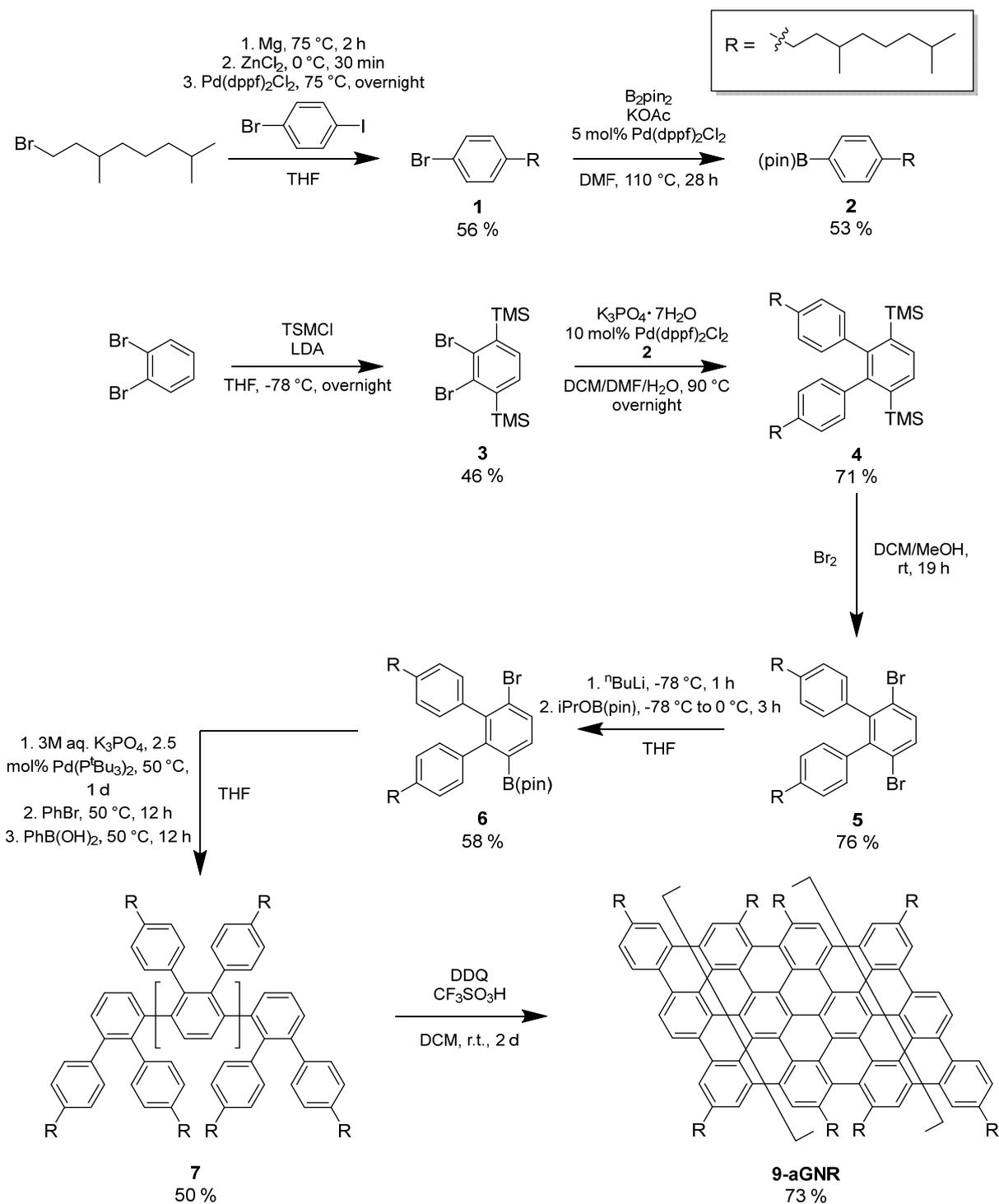

**Scheme S1. Synthesis of the 9-aGNR according to Li *et al.*[1]** Reaction conditions were kept constant. Yields of each reaction step are denoted under the respective molecule.

**1-bromo-4-(3,7-dimethyloctyl)benzene (1)**

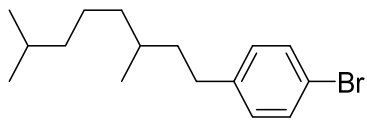

Magnesium shavings (1.47 g, 63.6 mmol) were added to a solution of 1-bromo-3,7-dimethyloctane (4.81 g, 21.7 mmol) in 10 mL dry, degassed THF and stirred at 70 °C for 2 h. The Grignard reagent was slowly added to a solution of dried $ZnCl_2$ (2.89 g, 21.2 mmol) in 20 mL of dry, degassed THF at 0 °C and stirred for 30 min. After addition of 1-bromo-4-iodobenzene (5.00 g, 17.7 mmol) and Pd(dppf)Cl$_2$ (388 mg, 0.53 mmol) the reaction mixture was heated to reflux overnight. When heating up the reaction mixture, addition of more THF was necessary as the reaction mixture used to solidify in the reaction vessel. After cooling to room temperature, 50 mL of diethyl ether and 50 mL of 1 M HCl were added to the reaction mixture. The organic phase was separated and the aqueous phase was extracted with 3 x 100 mL of diethyl ether. The combined organic phases were washed with brine and water, dried over $Na_2SO_4$ and the solvent was removed by rotary evaporation. The crude product was purified by flash column chromatography ($R_f$ = 0.9, pure petroleum ether (PE)) to yield **1** as a colorless oil (3.11 g, 59 %).

$^1$H NMR (CDCl$_3$, 300 MHz): δ 7.38 (m, 2H, Ar-H), 7.05 (m, 2H, Ar-H), 2.65-2.46 (m, 2H), 1.66-1.06 (m, 10H), 0.91 (d, J = 6 Hz, 3H, CH$_3$), 0.86 (d, J = 7 Hz, 6H, CH$_3$) ppm.

$^{13}$C NMR (CDCl$_3$, 300 MHz): δ 142.32, 131.46, 130.31, 119.38, 39.49, 38.99, 37.29, 33.09, 32.59, 28.15, 24.86, 22.89, 22.80, 19.75 ppm.

**2-(4-(3,7-dimethyloctyl)phenyl)-4,4,5,5-tetramethyl-1,3,2-dioxaborolane (2)**

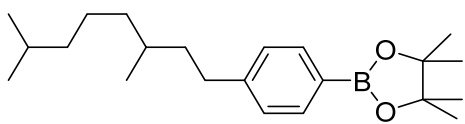

1-bromo-4-(3,7-dimethyloctyl)benzene (5.00 g, 16.8 mmol), 4,4,4',4',5,5,5',5'octamethyl-2,2'-bi-(1,3,2-dioxaborolane) (4.70 g, 18.5 mmol), potassium acetate (4.95 g, 50.5 mmol) and Pd(dppf)₂Cl₂ complex with DCM (dichloromethane, 687 mg, 0.84 mmol) were added to 35 mL of degassed, dry DMF and stirred at 110 °C for 28 h. After cooling down to room temperature 100 mL water and 100 mL ethyl acetate were added to the reaction mixture. The organic phase was separated and the aqueous phase was extracted with 3 x 100 mL of diethyl ether. The combined organic phases were washed with brine and water, dried over Na₂SO₄ and the solvent was removed under reduced pressure. The crude product was purified by flash column chromatography ($R_f$ = 0.4, DCM:PE 1:4) to yield **2** as a yellow oil (3.24 g, 56 %).

¹H NMR (CDCl₃, 300 MHz): δ 7.72 (m, 2H, Ar-H), 7.20 (m, 2H, Ar-H), 2.72-2.50 (m, 2H), 1.70-1.04 (m, 10H), 1.34 (s, 12H, CH₃) 0.91 (d, J = 6 Hz, 3H, CH₃), 0.86 (d, J = 7 Hz, 6H, CH₃) ppm.

**(2,3-dibromo-1,4-phenylene)bis(trimethylsilane) (3)**

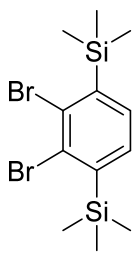

A mixture of dry, degassed THF (3.77 mL, 46.5 mmol), hexane (12.14 mL, 93.0 mmol) and toluene (4.92 mL, 46.5 mmol) was slowly added to solid lithium diisopropylamide (4.98 g, 46.5 mmol) at -78 °C. The slightly yellow LDA solution was then added dropwise to a solution of 1,2-dibromobenzene (5.00 g, 19.7 mmol) and trimethylsilylchloride (5.05 g, 46.5 mmol) at -78 °C. The reaction was stirred at -78 °C overnight. The reaction mixture was then hydrolysed with 50 mL of 0.1M H₂SO₄. The yellow organic phase was separated and the aqueous phase was extracted with 3 x 100 mL of diethyl ether. The combined organic phases were washed with brine and water, dried over Na₂SO₄ and the solvent was removed under reduced pressure. The crude product was purified by flash column chromatography ($R_f$ = 0.95, pure PE) and recrystallized from a 1:1 mixture of acetone and methanol to yield **3** as colorless crystals (3.43 g, 46 %).

¹H NMR (CDCl₃, 600 MHz): δ 7.33 (s, 2H, Ar-H), 0.39 (s, 18H, CH₃) ppm.

¹³C NMR (CDCl₃, 150 MHz): δ 145.93, 134.10, 133.52, -0.24 ppm.

### (4,4''-bis(3,7-dimethyloctyl)-[1,1':2',1''-terphenyl]-3',6'-diyl)bis(trimethylsilane) (4)

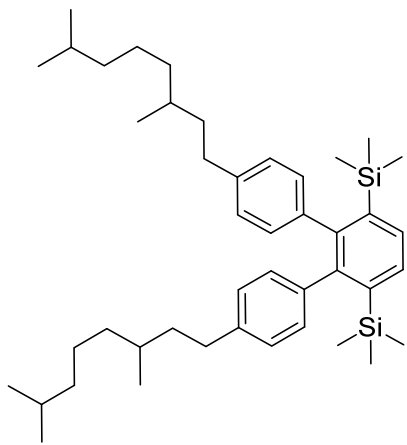

**2** (5.16 g, 15.0 mmol), **3** (1.90 g, 5.00 mmol), Pd(dppf)$_2$Cl$_2$ complex with DCM (408 mg, 0.50 mmol) and K$_3$PO$_4$ · 7 H$_2$O (6.36 g, 30.0 mmol) were dissolved in a mixture of 15 mL degassed DMF and 4 mL of degassed H$_2$O. The reaction mixture was stirred at 90 °C overnight. After cooling down to room temperature, 100 mL of ethyl acetate and 50 mL of water were added. The organic phase was separated and the aqueous phase was extracted with 3 x 100 mL of diethyl ether. The combined organic phases were washed with brine and water, dried over Na$_2$SO$_4$ and the solvent was removed by rotary evaporation. The crude product was purified by flash column chromatography (R$_f$ = 0.65, pure PE) to yield **4** as a slightly yellow oil (2.30 g, 69 %).

$^1$H NMR (CDCl$_3$, 400 MHz): δ 7.61 (s, 2H, Ar-H), 6.87-6.81 (m, 8H, Ar-H), 2.55-2.41 (m, 4H, CH$_2$), 1.56-1.06 (m, 20H, CH and CH$_2$), 0.88-0.85 (m, 18H, CH$_3$), -0.07 (s, 18H, CH$_3$) ppm.

$^{13}$C NMR (CDCl$_3$, 100 MHz): δ 147.74, 140.91, 140.29, 139.80, 132.66, 131.01, 130.98, 126.86, 126.82, 39.54, 39.06, 37.33, 33.20, 32.29, 28.16, 24.93, 22.90, 22.82, 19.78, 0.66 ppm.

**3',6'-dibromo-4,4''-bis(3,7-dimethyloctyl)-1,1':2',1''-terphenyl (5)**

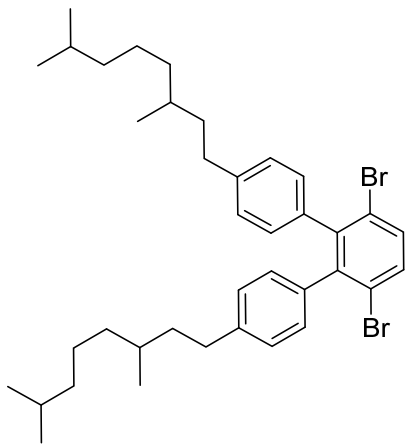

A solution of bromine (1.43 g, 0.46 mL, 8.94 mmol) in 2mL of DCM was slowly added to a solution of **4** (1.70 g, 2.56 mmol) in 3 mL of dry, degassed DCM and 5 mL of MeOH, which was held at 0 °C. The mixture was then stirred at room temperature overnight. After addition of 50 mL aqueous sodium sulphite solution and 50 mL of DCM, the organic phase was separated and the aqueous phase was extracted with 3 x 100 mL DCM. The combined organic phases were washed with brine and water, dried over $Na_2SO_4$ and the solvent was removed by rotary evaporation. The crude product was purified by flash column chromatography ($R_f$ = 0.35, pure PE) to yield **5** as a yellow oil (1.33 g, 78 %).

$^1$H NMR ($CDCl_3$, 400 MHz): δ 7.51 (s, 2H, Ar-H), 6.96-6.93 (m, 4H, Ar-H), 6.86-6.82 (m, 4H, Ar-H), 2.57-2.42 (m, 4H, $CH_2$), 1.58-1.06 (m, 20H, CH and $CH_2$), 0.88-0.86 (m, 18H, $CH_3$) ppm.

$^{13}$C NMR ($CDCl_3$, 100 MHz): δ 144.39, 142.02, 137.53, 132.71, 129.90, 129.88, 127.53, 123.60, 39.54, 38.76, 37.32, 33.32, 32.58, 28.17, 24.90, 22.91, 22.83, 19.83 ppm.

**2-(6'-bromo-4,4''-bis(3,7-dimethyloctyl)-[1,1':2',1''-terphenyl]-3'-yl)-4,4,5,5-tetramethyl-1,3,2-dioxaborolane (6)**

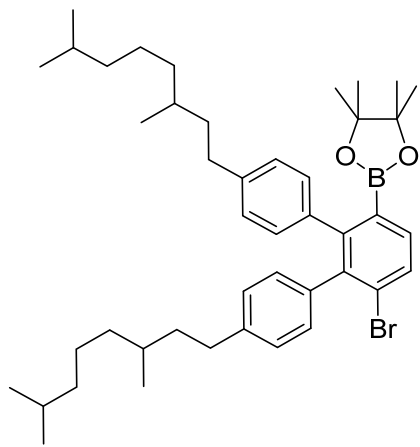

A 2.5 M n-Butyllithium solution in hexane (0.89 mL, 2.22 mmol) was added dropwise to a solution of **5** (1.35 g, 2.02 mmol) in 7 mL dry, degassed THF at -78 °C. The yellow reaction mixture was stirred at -78 °C for 1 h after which 2-isopropoxy-4,4,5,5-tetramethyl-1,3,2-dioxaborolane (0.52 g, 0.57 mL, 2.80 mmol) was added. The white reaction mixture was then stirred for 3 h, during which it was allowed to warm up to 0 °C. After this 50 mL of water and 50 mL of ethyl acetate were added. The organic phase was separated and the aqueous phase was extracted with 3 x 100 mL DCM. The combined organic phases were washed with brine and water, dried over $Na_2SO_4$ and the solvent was removed by rotary evaporation. The crude product was purified by flash column chromatography ($R_f$ = 0.3, 1:4 DCM:PE) to yield **5** as colorless oil (0.83 g, 57 %).

$^1$H NMR (CDCl$_3$, 400 MHz): δ 7.64 (d, J = 8 Hz, 1H, Ar-H), 7.44 (d, J = 8 Hz, 1H, Ar-H) 6.99-6.81 (m, 8H, Ar-H)), 2.64-2.43 (m, 4H, CH$_2$), 1.61-1.09 (m, 20H, CH and CH$_2$), 1.08 (s, 12H, CH$_3$), 0.91-0.81 (m, 18H, CH$_3$) ppm.

$^{13}$C NMR (CDCl$_3$, 100 MHz): δ 148.57, 141.74, 141.54, 141.15, 139.17, 137.70, 133.60, 131.02, 130.37, 130.33, 130.13, 127.45, 127.10, 126.81, 83.87, 39.58, 39.55, 39.30, 38.74, 37.35, 33.34, 33.24, 32.55, 32.42, 28.16, 24.91, 24.89, 24.68, 22.91, 22.89, 22.83, 22.79, 19.84, 19.73 ppm.

**Synthesis of precursor polymer (7)**

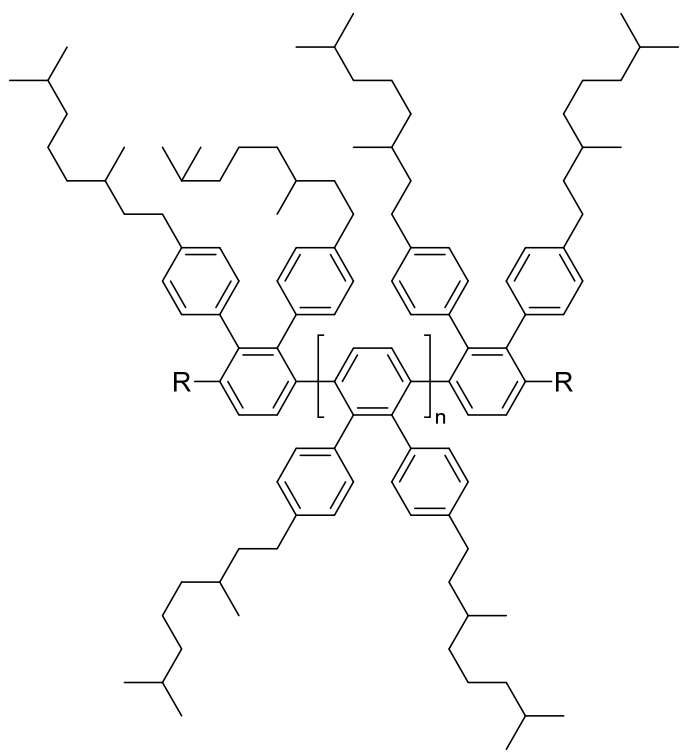

Inside of a nitrogen glovebox, the AB-type monomer **6** (102 mg, 0.14 mmol) and Pd(P$^t$Bu$_3$)$_2$ (4 mg, 2.5 mol%) were combined in a Schlenk flask. Under inert atmosphere, 3 mL of degassed THF and 0.7 mL of a 3M K$_3$PO$_4$ solution in degassed water were added to the reaction mixture. The mixture was stirred at 50 °C for 24h. After that, bromobenzene (21 mg, 0.14 mmol) was added and the reaction was stirred at 50 °C for 8 h. Finally, phenyboronic acid (17 mg, 0.14 mmol) was added and the reaction was stirred at 50 °C for another 8 h. After cooling down to room temperature 10 mL of water and 10 mL of DCM were added to the reaction mixture. The organic phase was separated and the aqueous phase was extracted with 3 x 10 mL DCM. The volume of the combined organic phases was reduced to ~2 mL under reduced pressure and the resulting solution was added to 20 mL of MeOH at room temperature. During this process the polymer precipitates as a white solid. After stirring at room temperature for 1 h, the polymer was filtered and dried under vacuum overnight. To remove impurities and smaller polymer strands, the crude polymer was washed by Soxhlet extraction with acetone for 2 days.

Note that the chemical nature of the end groups R was not determined here. According to Li *et al.* the synthesis usually results in H/H and sometimes H/Phenyl end groups.[1]

**Synthesis of 9-aGNR**

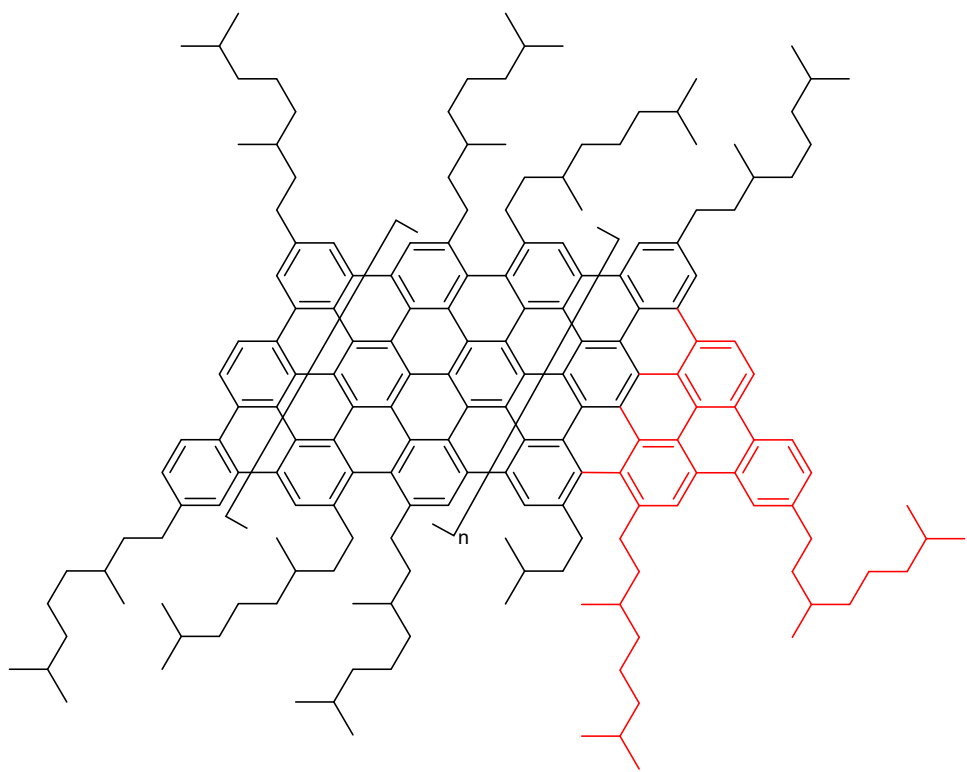

Precursor polymer **7** (202 mg) and DDQ (459 mg) were dissolved in dry, degassed DCM. Dropwise addition of triflic acid (8.6 mL) at 0 °C led to a color change of the reaction mixture from yellow to black. The reaction mixture was stirred at room temperature for 2 days. Subsequently, the reaction mixture was quenched with saturated $NaHCO_3$ solution until no more $CO_2$ emerged upon addition (after ~30 mL). The black precipitate was filtered off and washed thoroughly with 200 mL of water, methanol and acetone each. The crude product was then purified by Soxhlet extraction with acetone for 2 days.

Note that depending on the number of terphenyl units in the precursor polymer, the resulting 9-aGNR can either have a parallel shape (black structure, even number of terphenyl units) or a trapezoidal shape (black + red structure, odd number of terphenyl units).

# NMR spectra

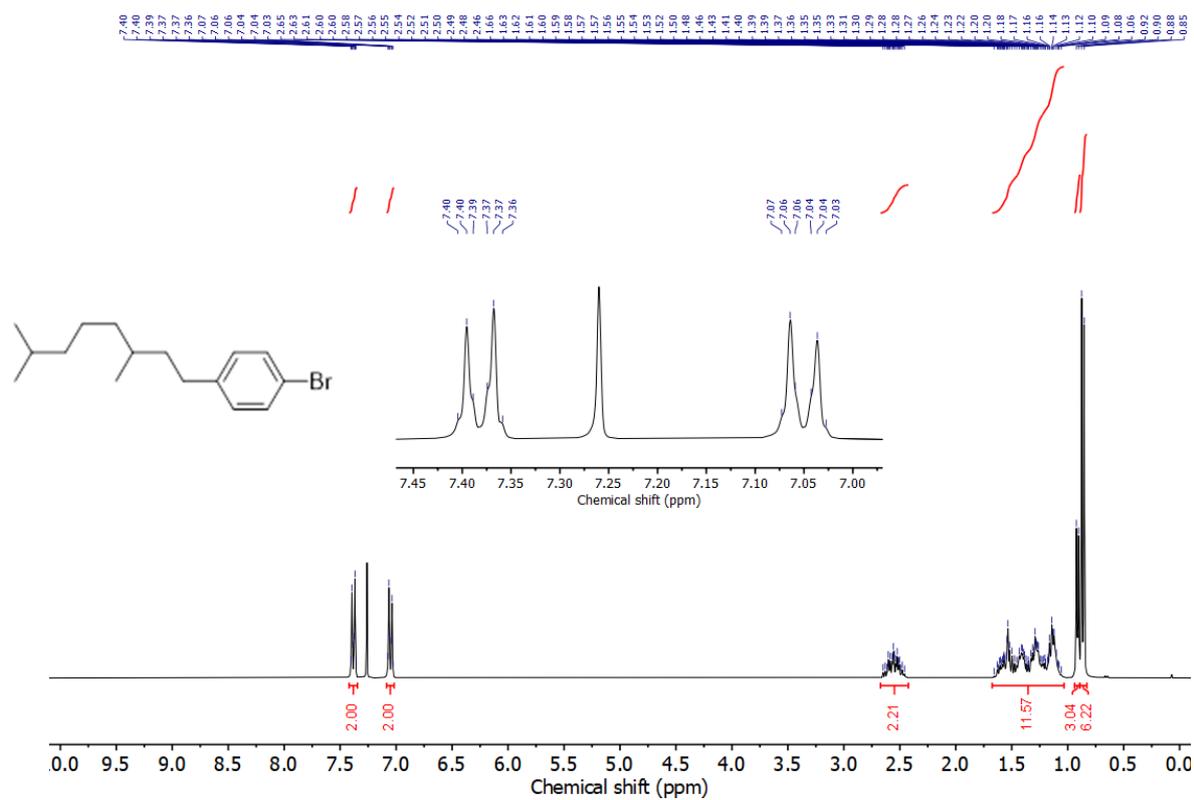

$^1$H NMR spectrum of **1** (CDCl$_3$, 300 MHz, 25°C)

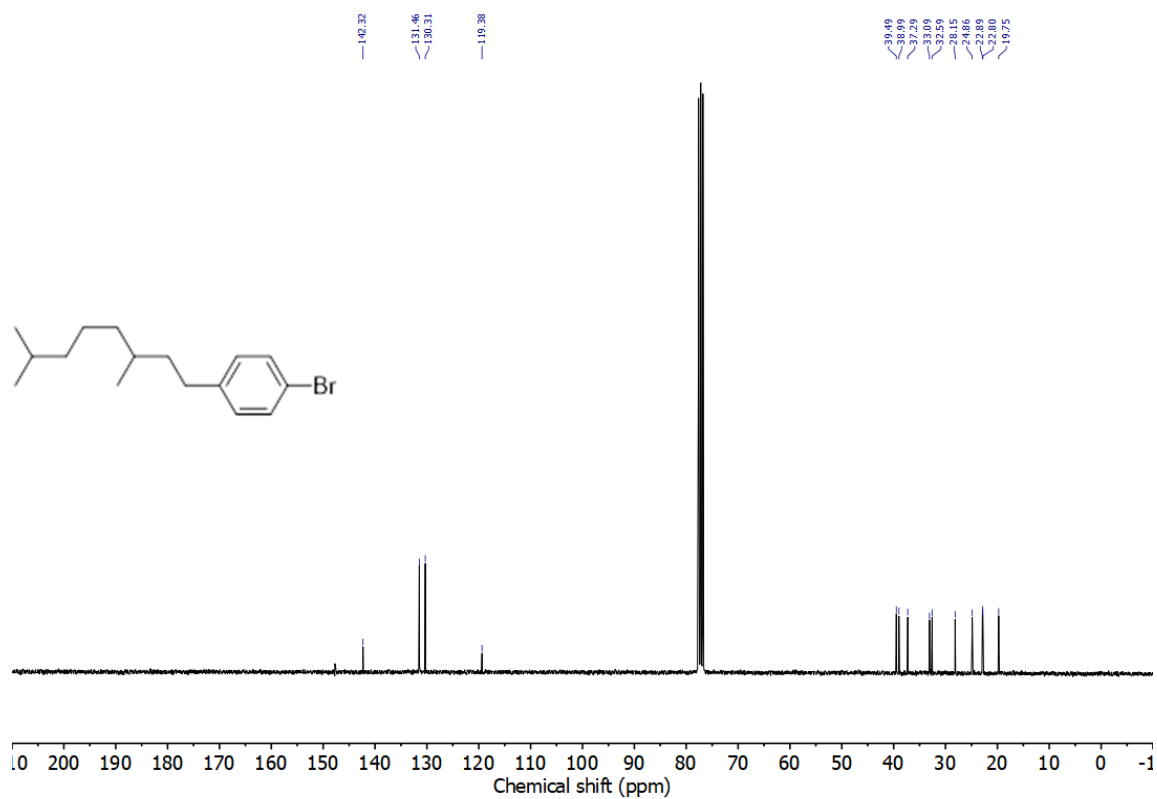

$^{13}$C NMR spectrum of **1** (CDCl$_3$, 75 MHz, 25°C)

S-12

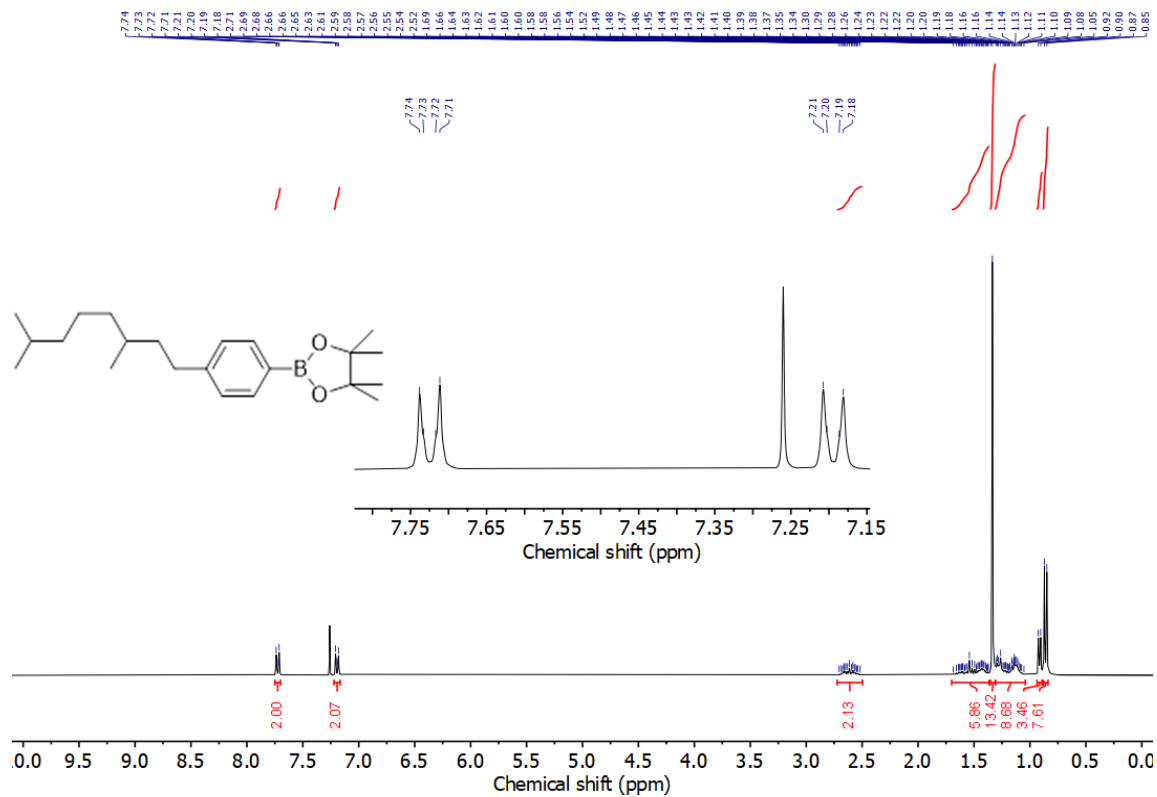

$^1$H NMR spectrum of **2** (CDCl$_3$, 300 MHz, 25°C)

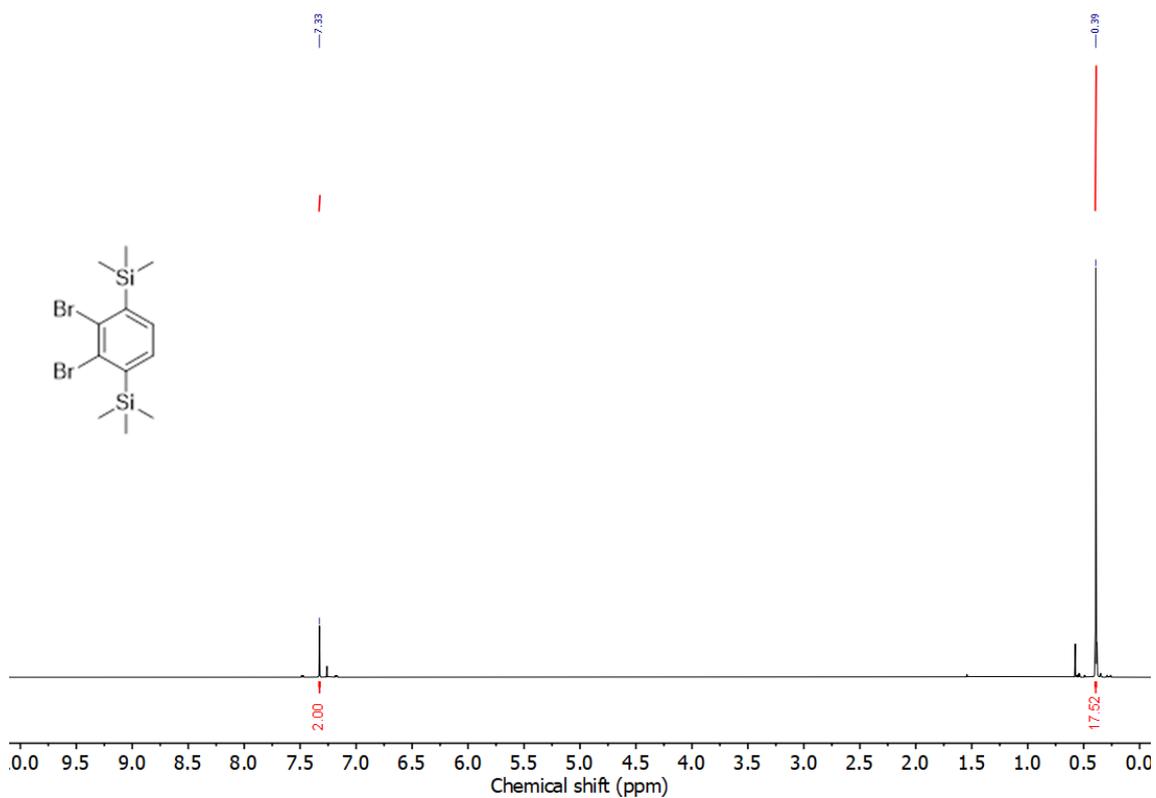

$^1$H NMR spectrum of **3** (CDCl$_3$, 600 MHz, 25°C)

S-13

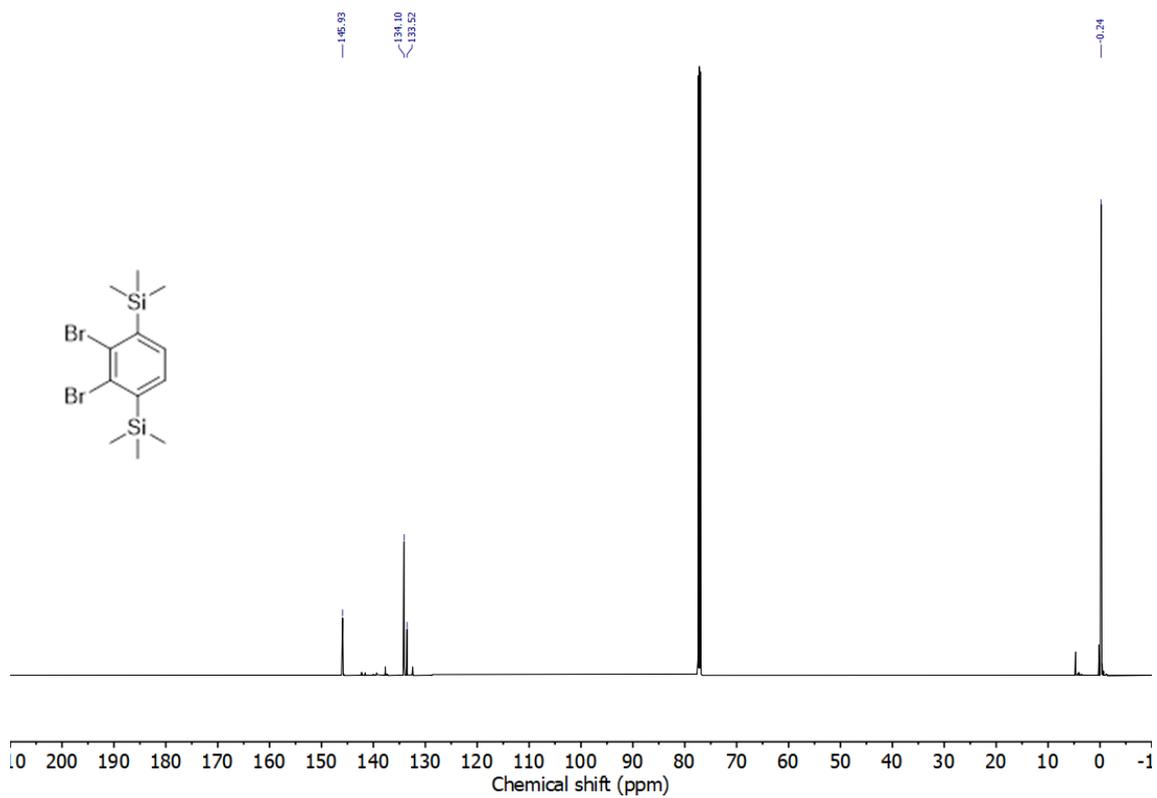

$^{13}$C NMR spectrum of **3** (CDCl$_3$, 150 MHz, 25°C)

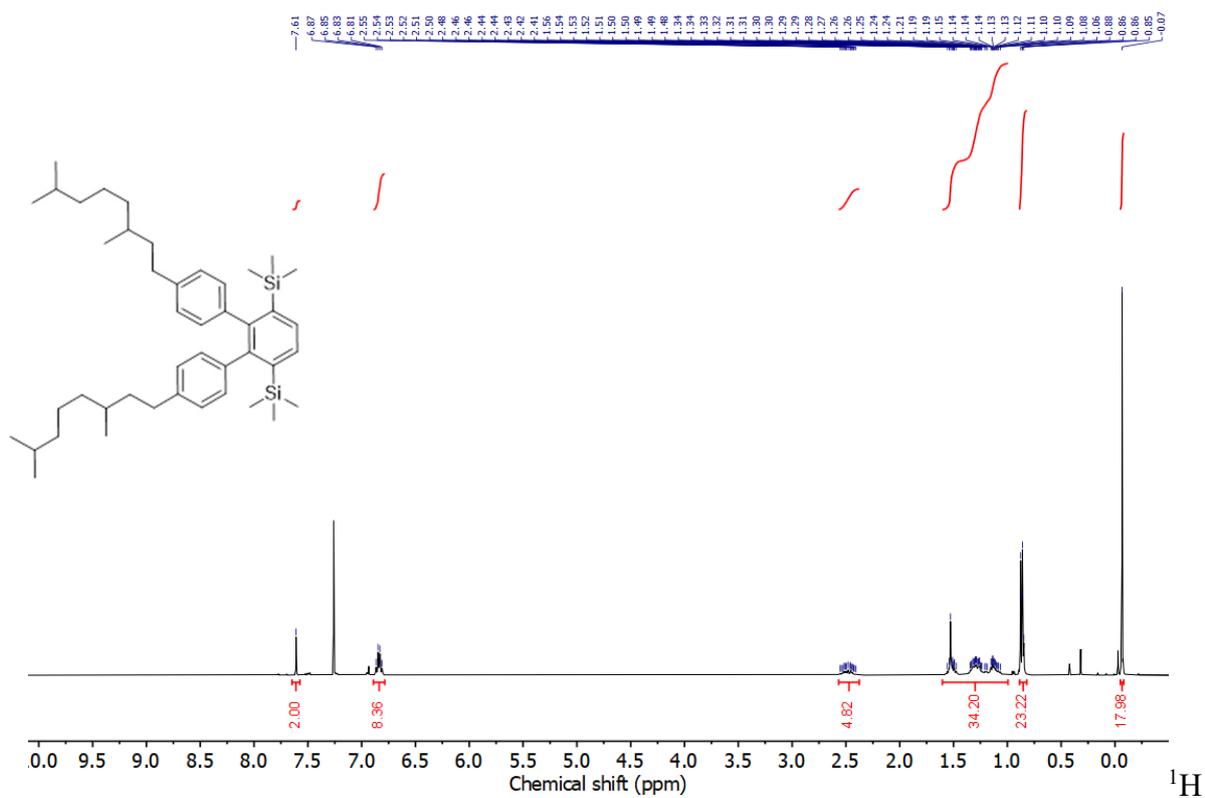

$^1$H NMR spectrum of **4** (CDCl$_3$, 400 MHz, 25°C)

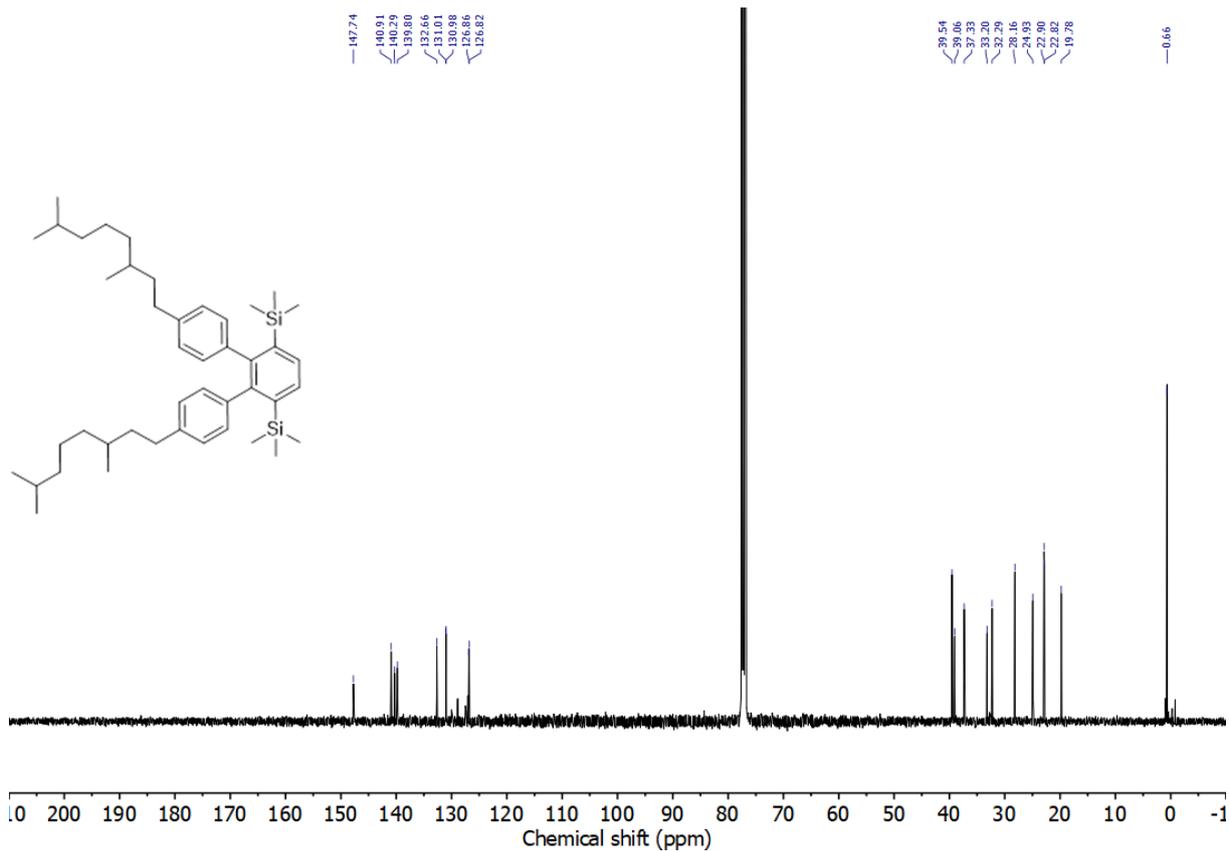

13C NMR spectrum of **4** (CDCl3, 100 MHz, 25°C)

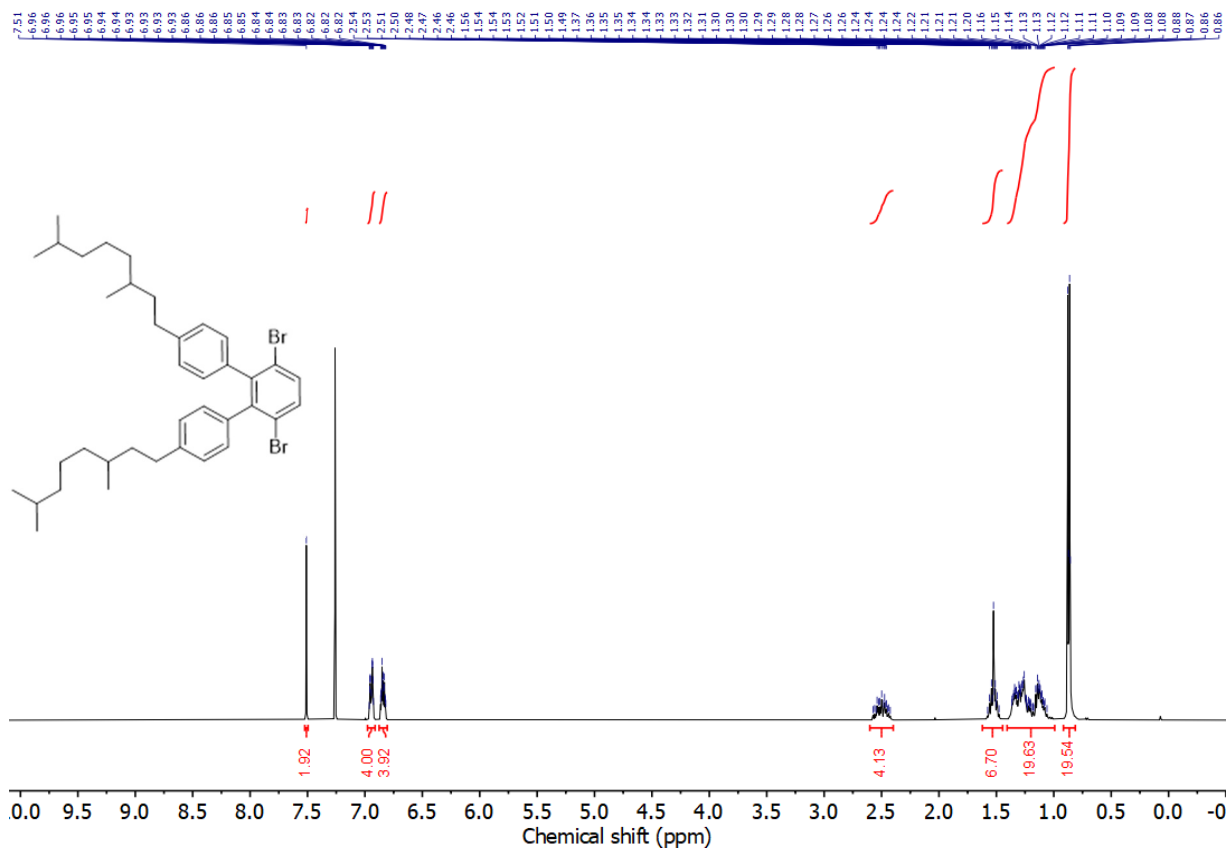

¹H NMR spectrum of **5** (CDCl₃, 400 MHz, 25°C)

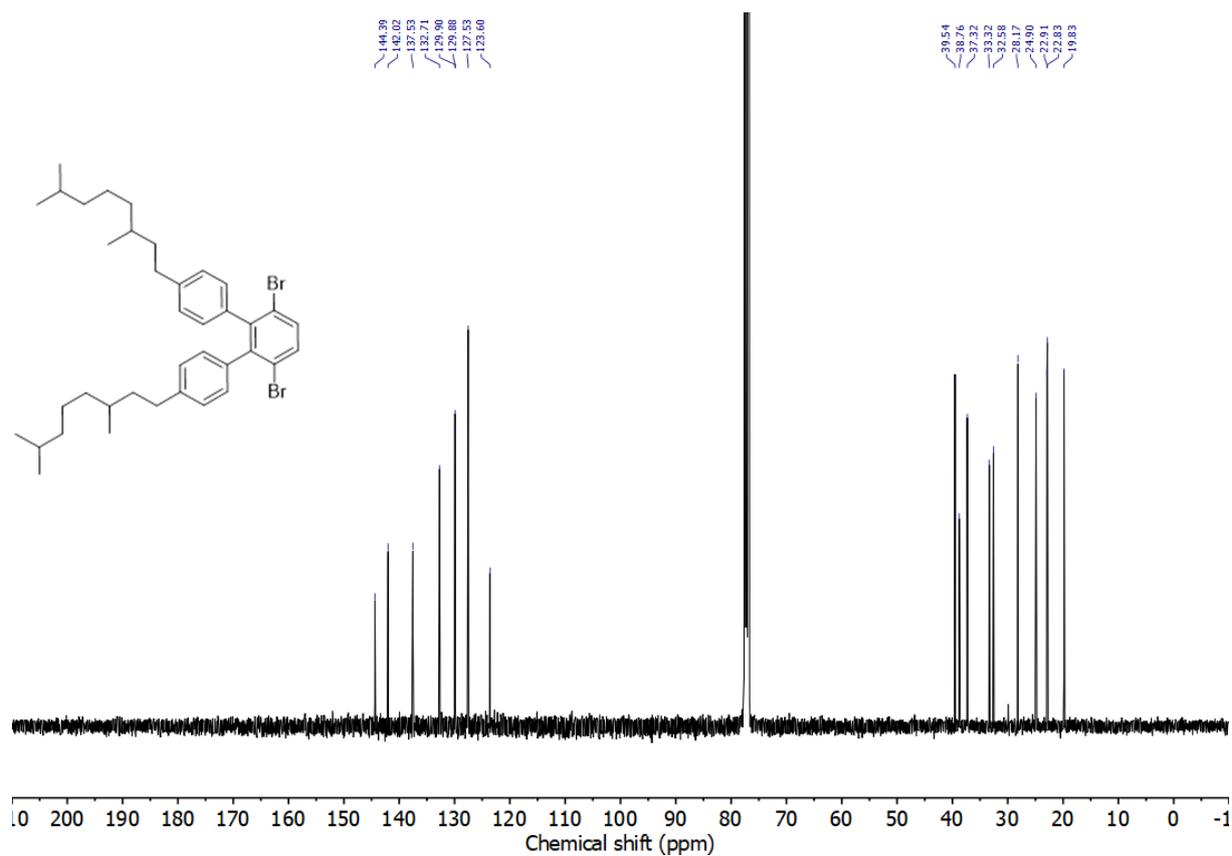

¹³C NMR spectrum of **5** (CDCl₃, 100 MHz, 25°C)

¹H NMR spectrum of **6** (CDCl₃, 400 MHz, 25°C)

¹³C NMR spectrum of **6** (CDCl₃, 100 MHz, 25°C)

**Supporting Figures – Spectroscopic Characterization 9-aGNRs**

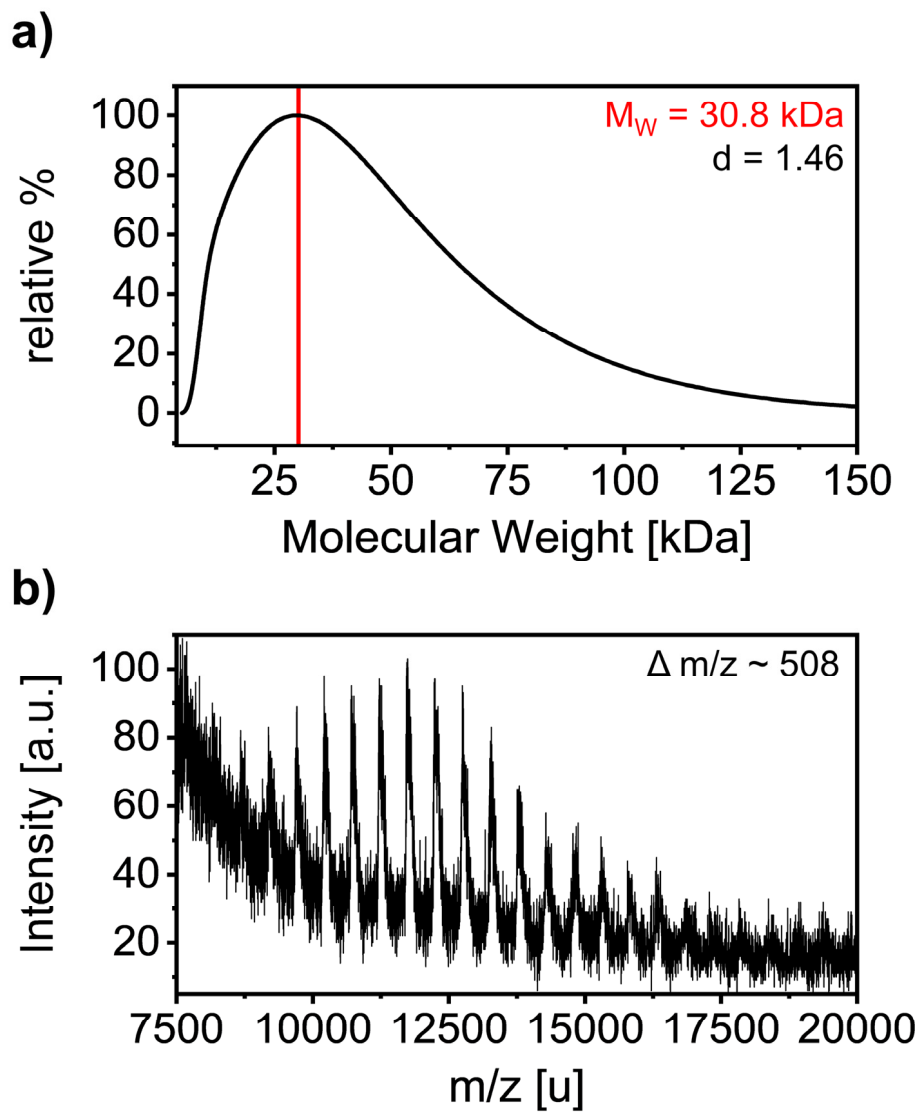

**Figure S1. Length Characterization of 9-aGNR precursor polymer. a)** Size exclusion chromatogram of a solution of precursor polymer in THF against a polystyrene standard. The determined $M_W$ is marked by a red line in the chromatogram. The depicted $M_W$ corresponds to 60 coupled terphenyl units and a length of ~25 nm. **b)** MALDI-TOF mass spectrum of precursor polymer. Peaks are spaced by ~508 u which corresponds to the molecular weight of a monomer unit. Peaks can be clearly observed up to a m/z ratio of 17500, indicating a length of ~35 terphenyl units, corresponding to ~15 nm.

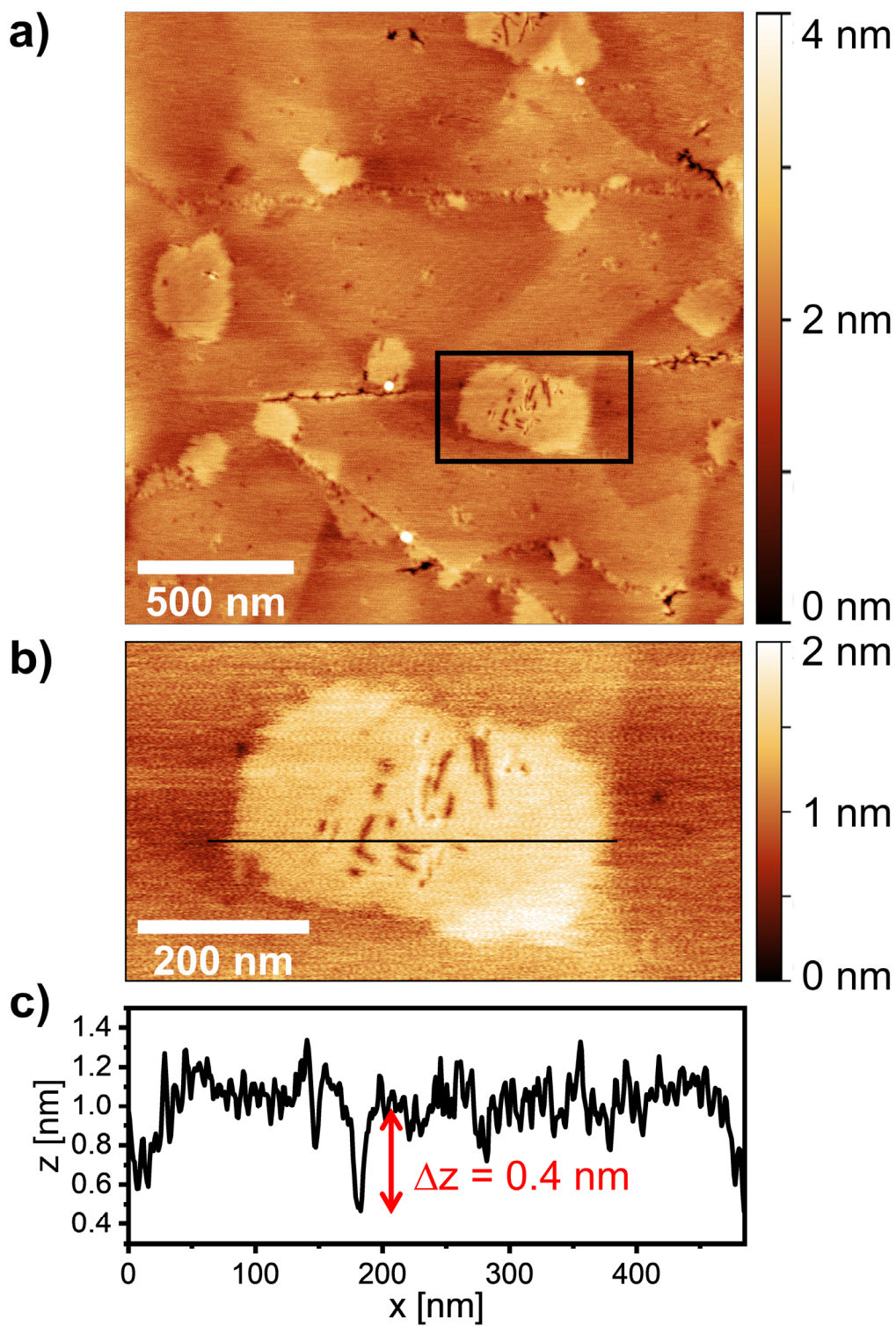

**Figure S2. a)** Tapping-mode AFM image (Bruker Dimension Icon equipped with OLTESPA-V3 tips) of self-assembled 9-aGNRs on freshly cleaved HOPG. 15 μL of a diluted dispersion (optical density of 0.04 at 820 nm) of the 0.2-1 k*g* fraction 9-aGNR in toluene were drop-cast onto a freshly cleaved HOPG substrate. The substrate temperature was held at 50 °C for 5 minutes and then increased to 80 °C for another 10 min to facilitate slow evaporation of solvent. The image shows several small irregular islands that originate from GNR self-assembly

S-19

**b)** Zoom-in on the highlighted area in (a) showing an island of self-assembled GNRs. While individual GNRs cannot be resolved, ribbon-shaped vacancies in the island indicate the presence of GNRs with different lengths in the dispersion. **c)** Height profile along the line in (b) confirming the height of a nanoribbon island as 0.4 nm as previously reported by Li *et al.* [1]

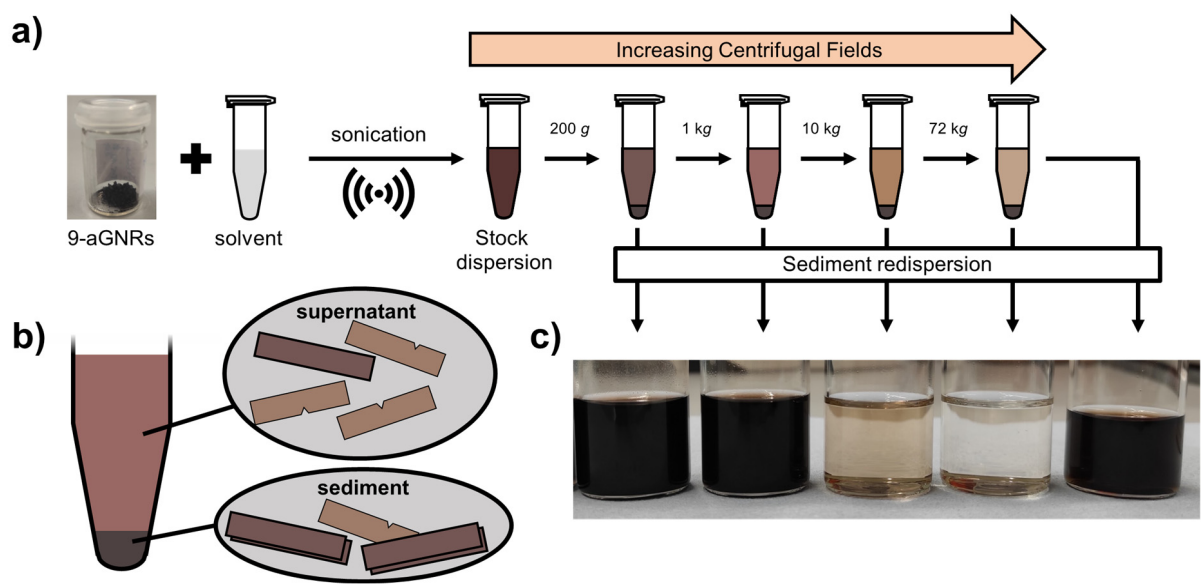

**Figure S3. a)** Schematic depiction of dispersion and liquid cascade centrifugation (LCC) process. 9-aGNRs are exfoliated in THF or toluene *via* sonication. The resulting stock dispersion is then centrifuged at 200 *g*. The supernatant is carefully removed and subjected to another centrifugation step at higher RCF, while the sediment is redispersed in fresh solvent. This process is repeated at RCF values of 1 k*g*, 10 k*g* and 72 k*g*. **b)** The sediment is expected to contain aggregated and mostly defect-free GNRS, while the supernatant should contain mostly defective GNRs. With each centrifugation step, more defect-free GNRs are removed from the dispersion. **c)** Photograph of the resulting dispersions.

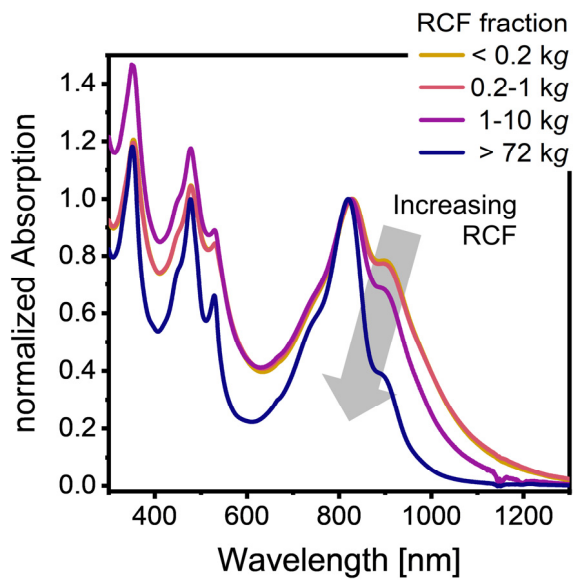

**Figure S4.** Absorption spectra for resulting 9-aGNR dispersions in toluene after LCC. A slight redshift of ~2 nm of the peak maxima compared to dispersions in THF can be explained with the different dielectric environment of the two solvents.

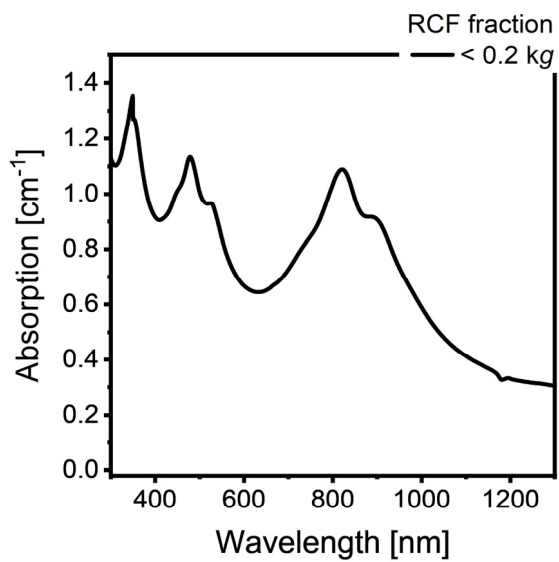

**Figure S5. Uncorrected absorption spectrum for <0.2 k*g* fraction in THF.** The spectrum shows a scattering background with an onset of ~0.3 at 1300 nm.

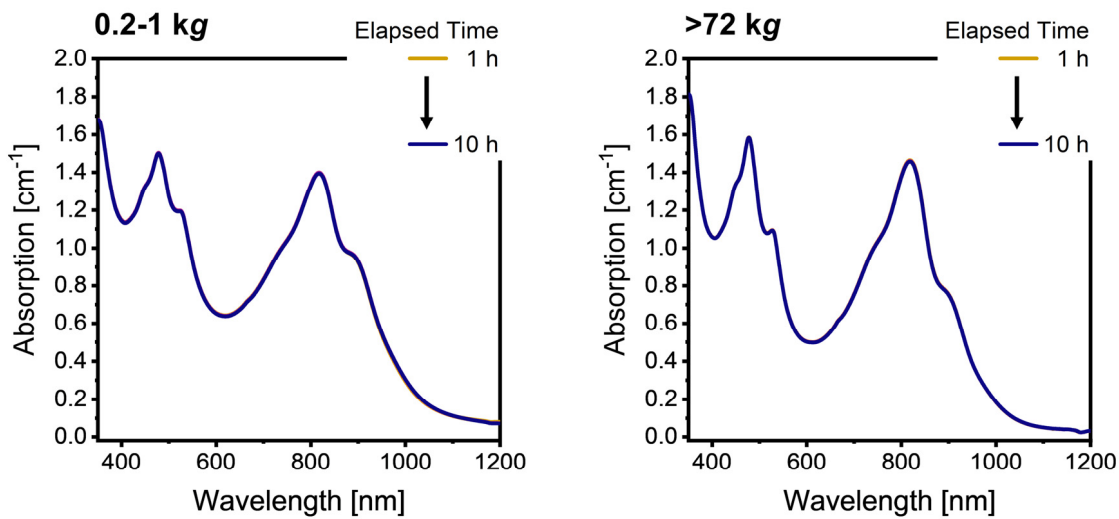

**Figure S6. Absorption spectra of exfoliated 9-aGNRs in THF over 10 h.** Spectra were measured once per hour for 10 h without moving the cuvette. As absorption spectra do not change, the first spectrum is barely visible. This indicates a high dispersion stability.

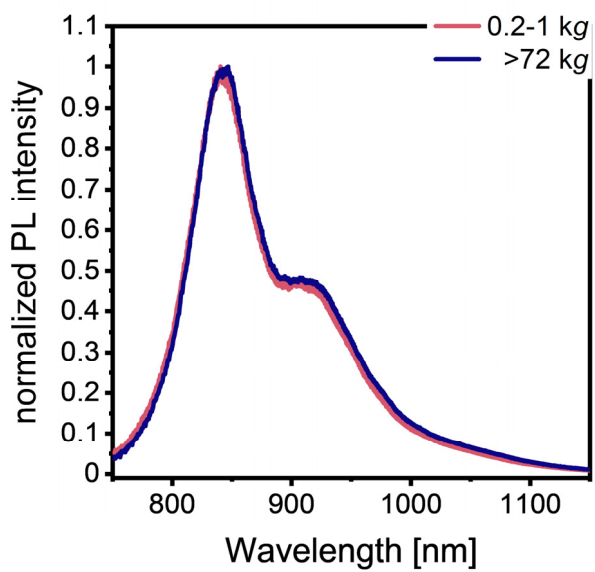

**Figure S7.** PL spectra of a 0.2-1 k*g* >72 k*g* 9-aGNR fraction in toluene.

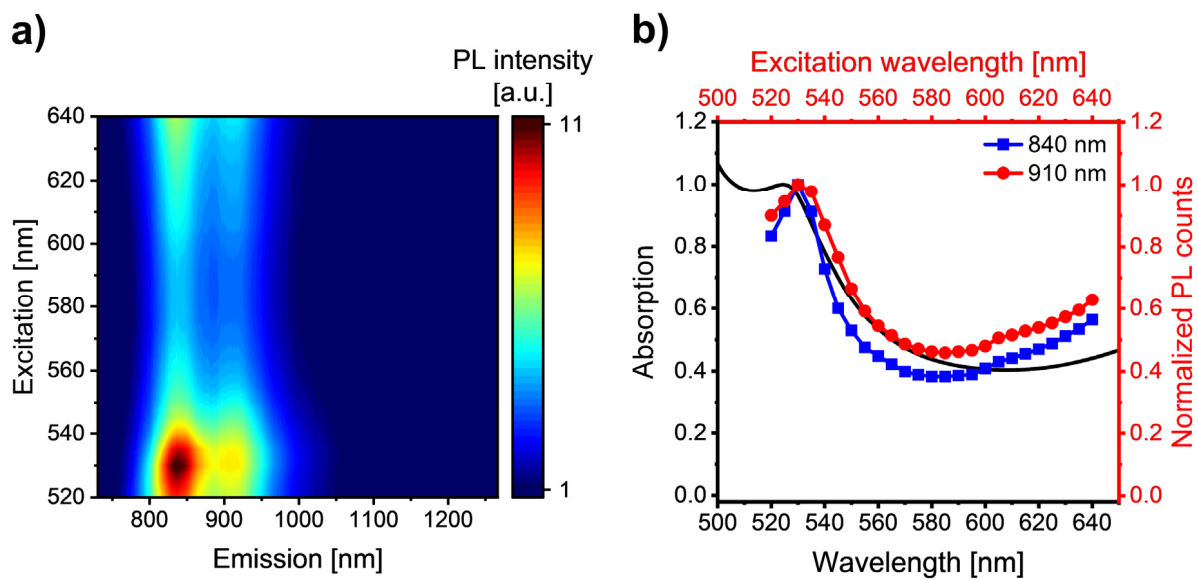

**Figure S8. a)** PL excitation-emission map of a >72 k*g* 9-aGNR fraction. **b)** Excitation spectrum obtained for emission at 840 nm (blue) and 920 nm (red). An absorption spectrum of the corresponding >72 k*g* fraction normalized to the absorption at 525 nm is shown as a black line for comparison.

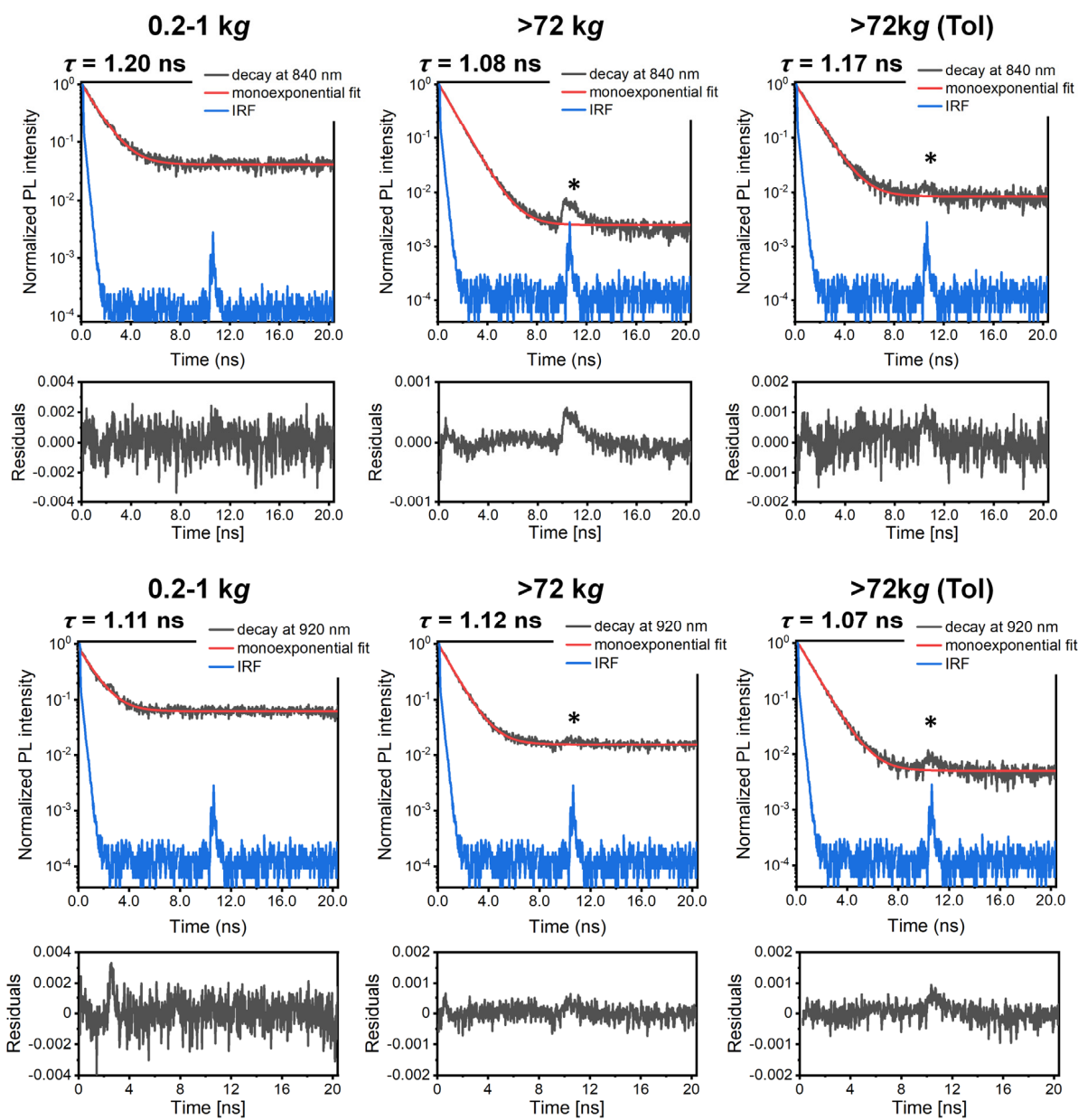

**Figure S9.** TCSPC histograms for PL decay at the emission features at 840 nm (upper panel) and 920 nm (lower panel) under excitation at 535 nm. The measured instrument response function (IRF) is shown in blue. Histograms were measured for 0.2-1 k*g* and >72 k*g* 9-aGNR fractions in THF and a >72 k*g* 9-aGNR fraction in toluene. The peak at 10 ns (marked with an asterisk) also appears in the IRF and is an artefact of the TCSPC setup. All histograms were fitted with a mono-exponential tail-fit procedure (fit function $f(t) = A \cdot e^{\frac{t-t_o}{\tau}} + c$, where $\tau$ is the extracted PL lifetime). Residuals of the fitting procedures are shown below. The extracted lifetimes vary between 1.0 and 1.2 ns with those at 920 nm being insignificantly shorter than those at 840 nm. Although fitting two exponential decays with different amplitudes should be possible for the expected two different GNR species, such multi-parameter fits would not be reliable for the available data.

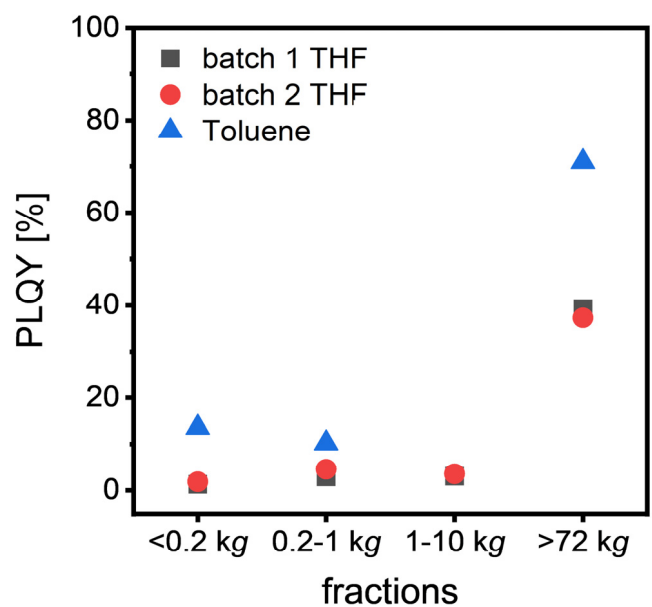

**Figure S10.** PL quantum yields (PLQY) of 9-aGNR dispersions in THF and toluene. PLQYs increase with increasing RCF.

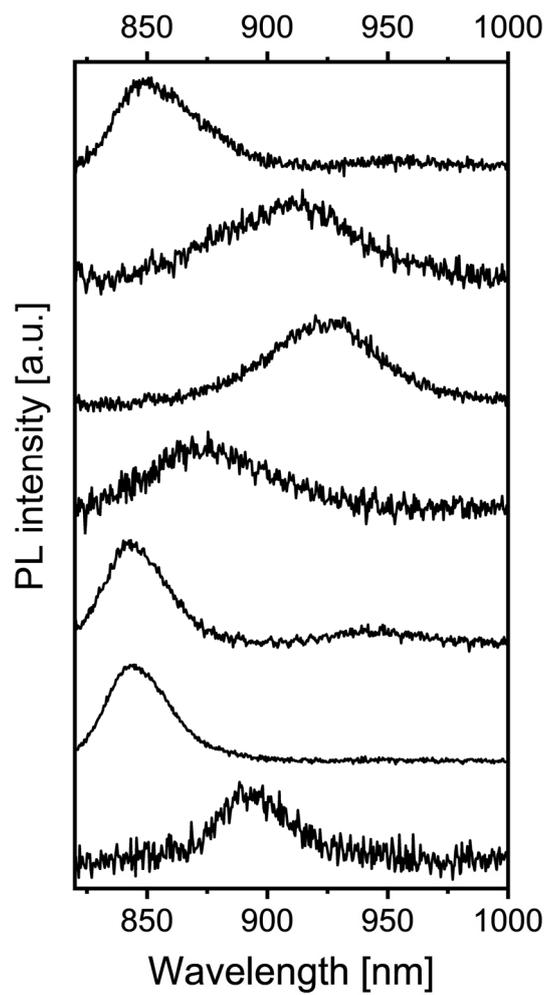

**Figure S11.** Room temperature PL spectra of 0.2-1 k*g* 9-aGNRs embedded in a polystyrene matrix.

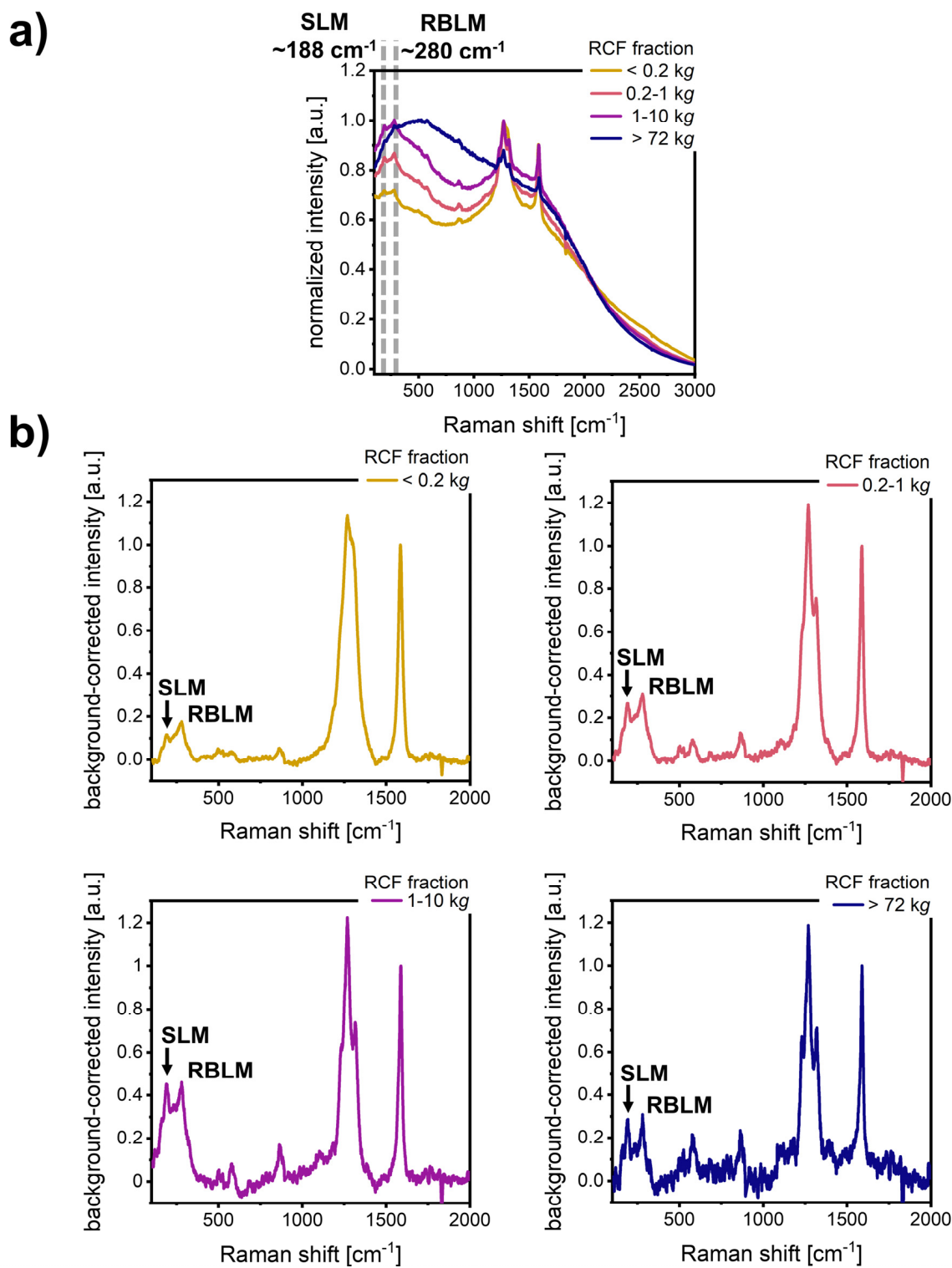

**Figure S12.** Raman spectra of exfoliated 9-aGNRs excited with a 785 nm laser. **a)** Uncorrected Raman spectra showing a high PL background. The RBLM at 280 cm$^{-1}$ and SLM at 188 cm$^{-1}$ (dashed lines) are clearly visible in all LCC fractions. **b)** Background-corrected Raman spectra for all fractions.

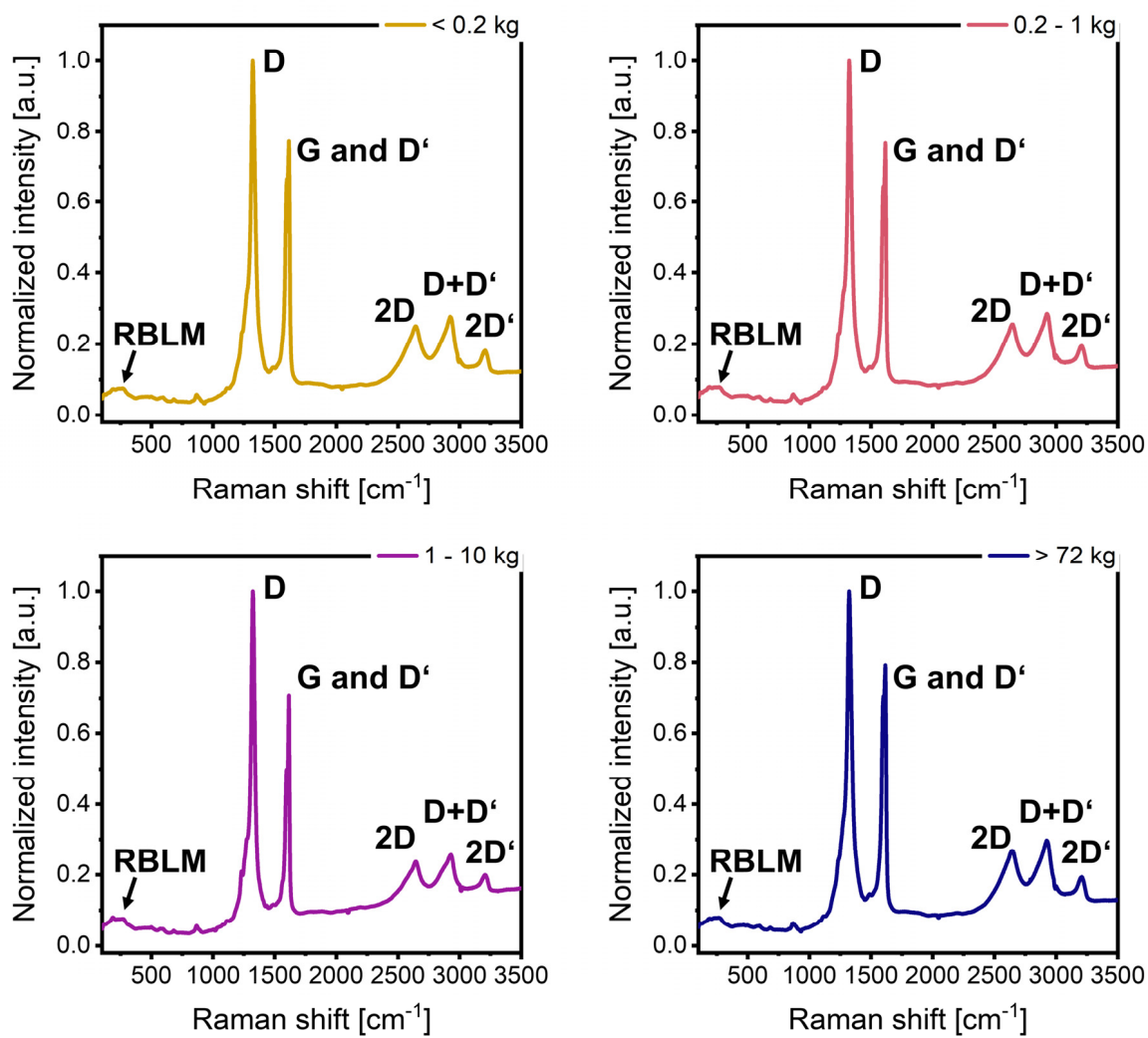

**Figure S13.** Full range Raman spectra of all LCC fraction of exfoliated 9-aGNRs in THF (as shown in Figure 4 of the main text) excited with a 532 nm laser.

# Quantum chemical calculations of neutral 9-aGNRs

All optimized geometries are available at the following data repository:

https://doi.org/10.11588/data/JAV0ZK

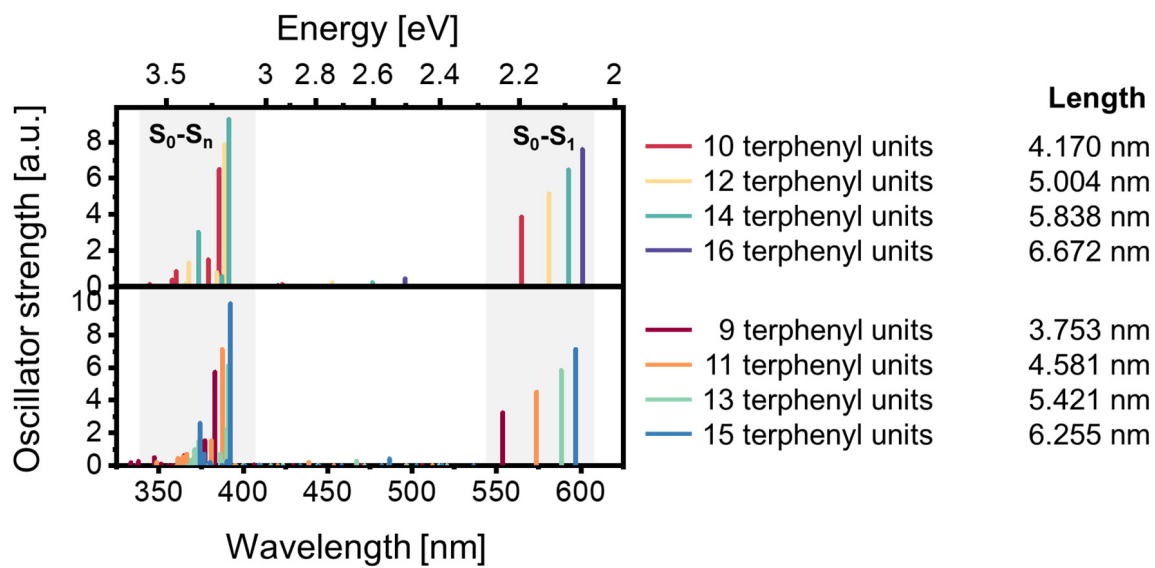

**Figure S14**: Comparison between the excited state vertical transition energies as computed at the TD-DFT (ωB97X-D/6-31G*) level for different oligomers with parallel (top) and trapezoidal (bottom) shapes.

**Top view** **Side view**

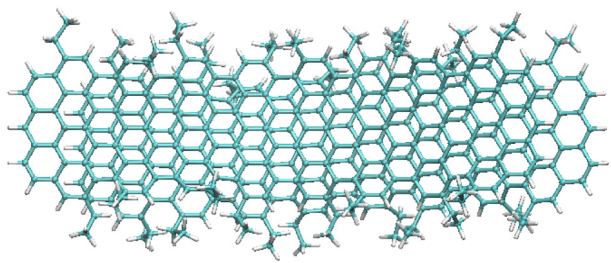
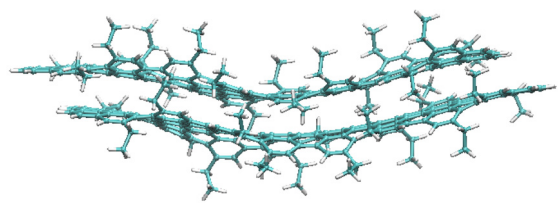
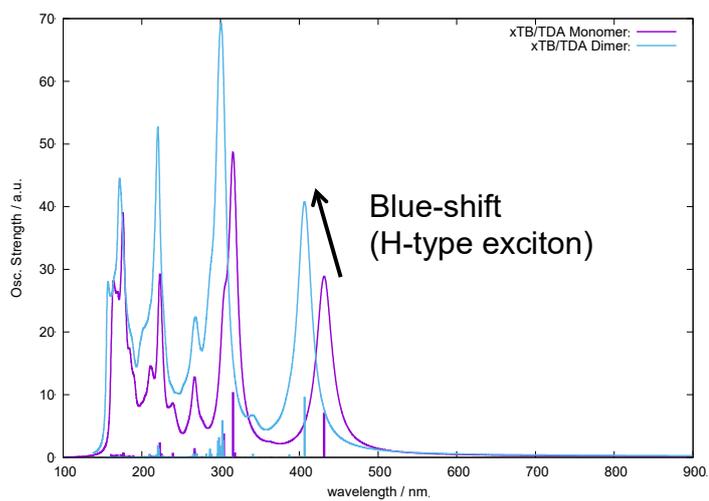

**Figure S15**. GFN2-xTB optimized geometry for the 9-aGNR dimer (top and side views). Comparison between the sTD-DFT transition energies (unscaled values) computed for the monomer (purple line) and dimer (light blue line).

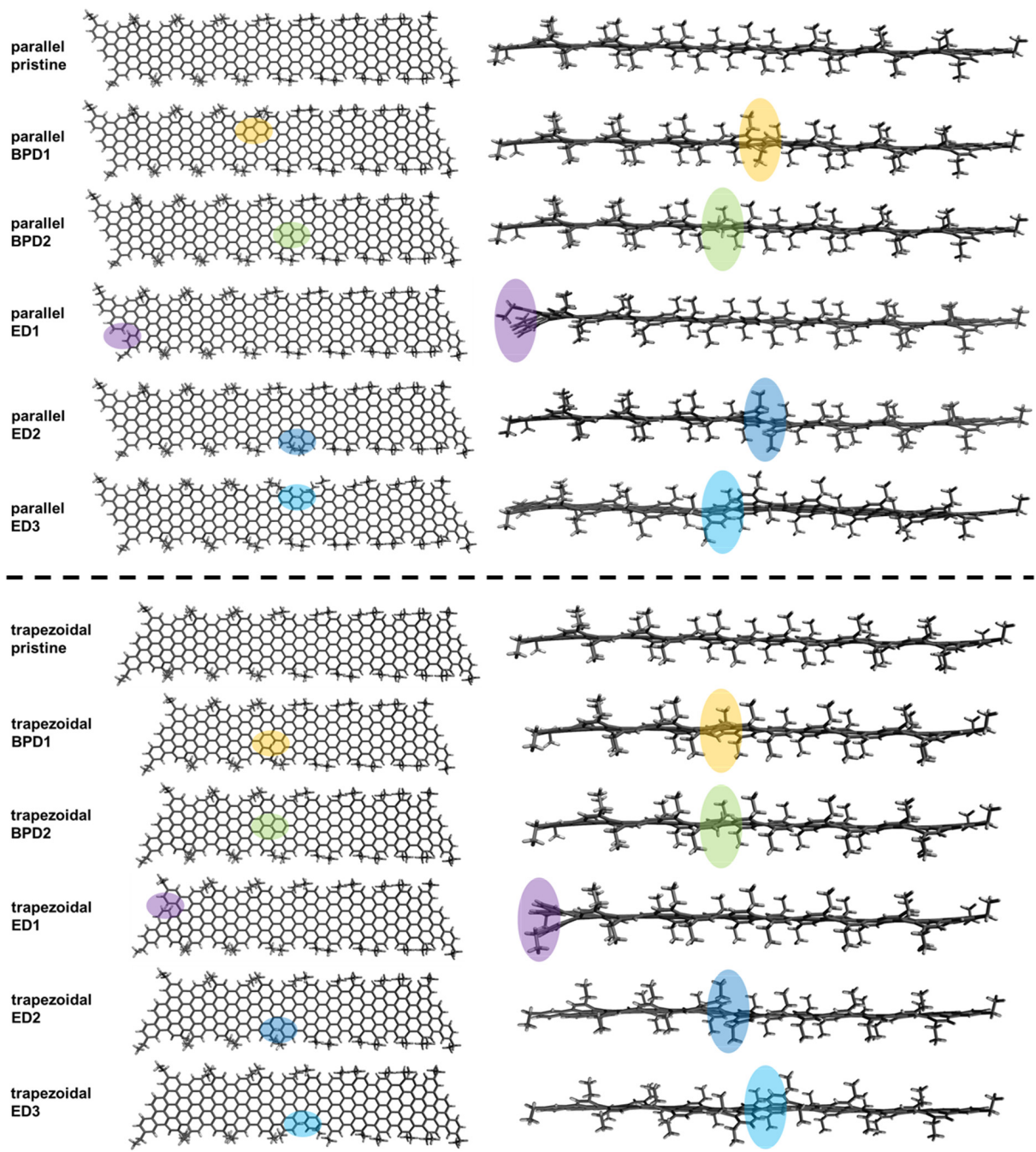

**Figure S16**. Optimized (GFN2-xTB) structures for parallel 9-aGNRs (top panel) and trapezoidal 9-aGNRs (bottom panel). The coloured regions highlight the positions of the respective defect.

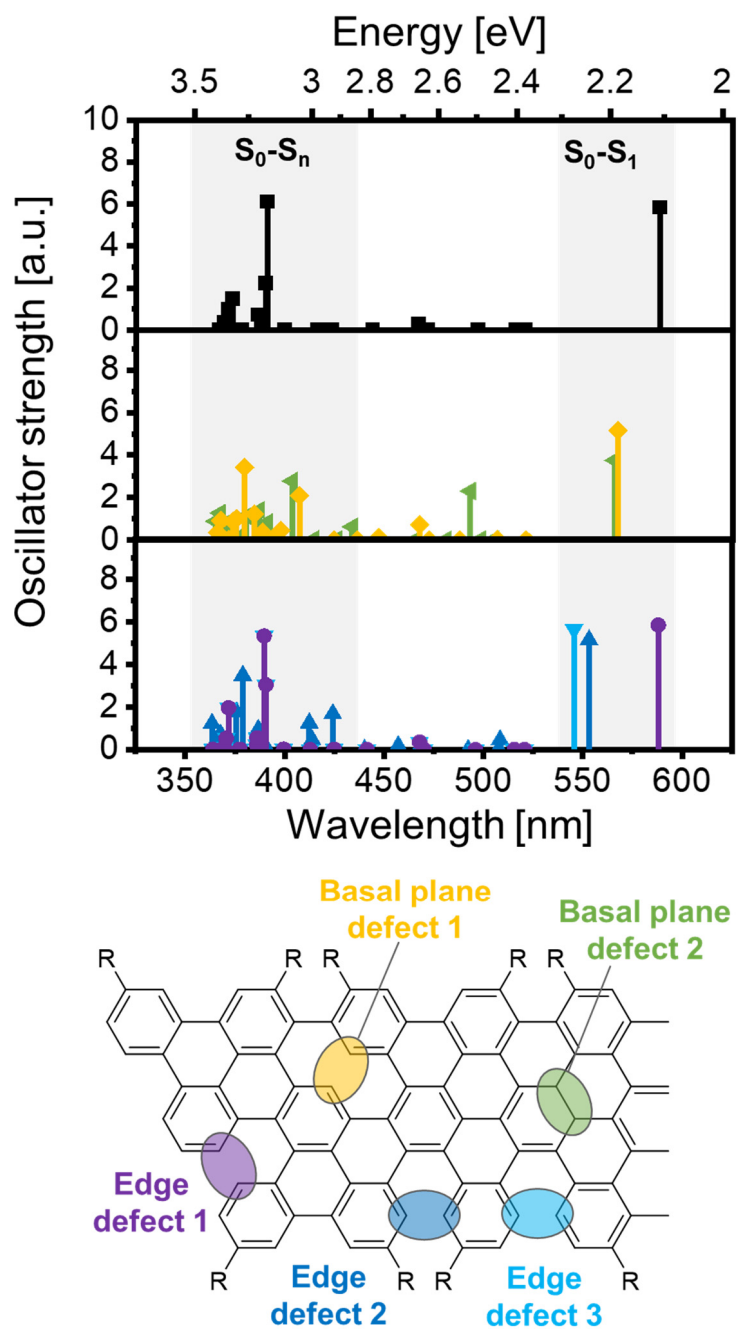

**Figure S17**: TD-DFT transition energies computed for the **trapezoidal 9-aGNR**: comparison between pristine and defected nanoribbons. Pristine (top panel, black line), BP1 (middle panel, yellow line), BP2 (middle panel, green line), ED1 (bottom panel, purple line), ED2 (bottom panel, dark blue line), ED1 (bottom panel, light blue line).

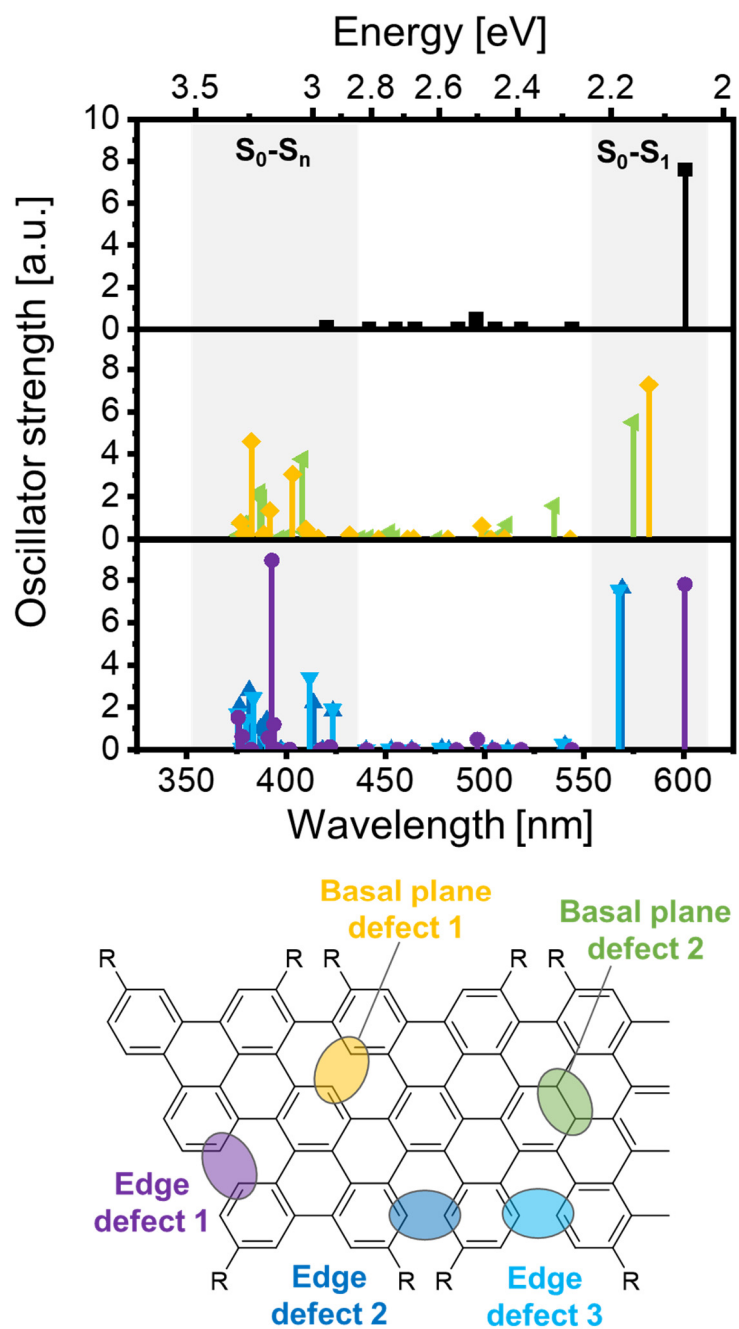

**Figure S18**: TD-DFT transition energies computed for the **parallel 9-aGNR**: comparison between pristine and defected nanoribbons. Pristine (top panel, black line), BP1 (middle panel, yellow line), BP2 (middle panel, green line), ED1 (bottom panel, purple line), ED2 (bottom panel, dark blue line), ED1 (bottom panel, light blue line).

## Supporting Figures – Doped 9-aGNRs

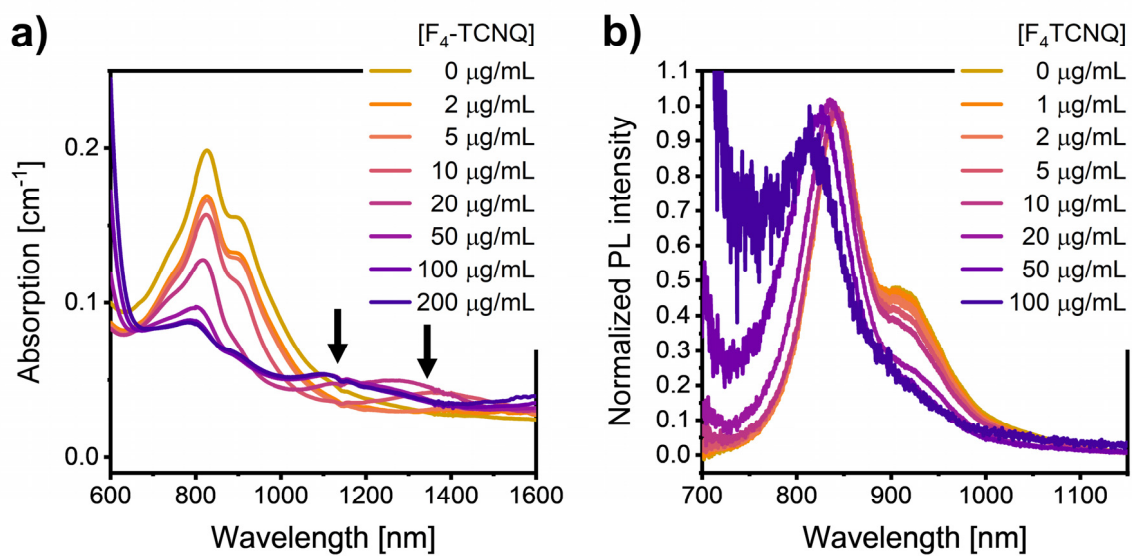

**Figure S19. a)** Absorption spectra of a >0.2-1 k*g* 9-aGNR dispersion doped with F₄TCNQ. Charge-induced absorption peaks are marked by arrows. **b)** Normalized PL spectra of a > 0.2-1 k*g* 9-aGNR dispersion doped with F₄TCNQ showing blue-shift and concentration-dependent quenching of the main emission peaks.

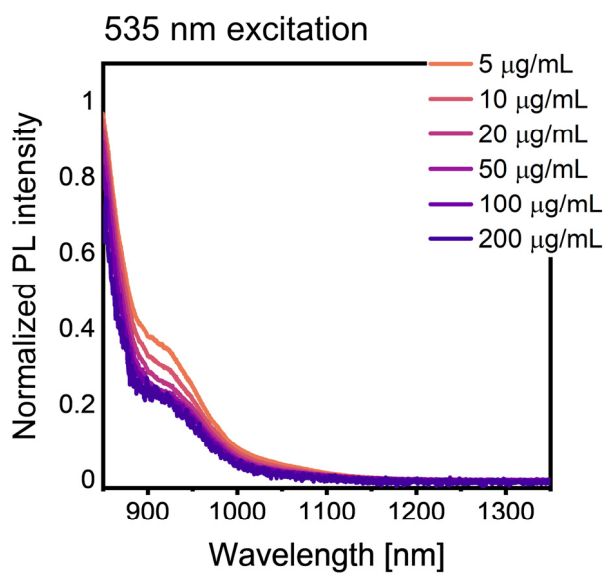

**Figure S20.** PL spectra of F₄TCNQ-doped >72 k*g* 9-aGNR dispersions recorded at longer wavelengths. No new red-shifted emission features are visible.

## Quantum chemical calculations of charged 9-aGNRs

The structure of doped (i.e., positively charged species, +1) 9-aGNRs was optimized at the GFN2-xTB level. The electronic transitions were computed at the TD-UDFT level (UωB97X-D/6-31G*).

The calculated electronic transitions of undoped - neutral (yellow bars) vs. doped - charged (purple bars) 9-aGNR of trapezoid (upper panel) and parallel (bottom panel) shapes are compared in **Figure S21.** For each species (in their neutral and charged state) we considered the pristine case (without defects) and the presence of an edge defect type 2 (ED2). Charged species show dipole allowed electronic transitions lower in energy than the neutral species (see 700-1000 nm region). The shapes (trapezoid or parallel) and the structural defects, slightly affect the transition energies of charged species.

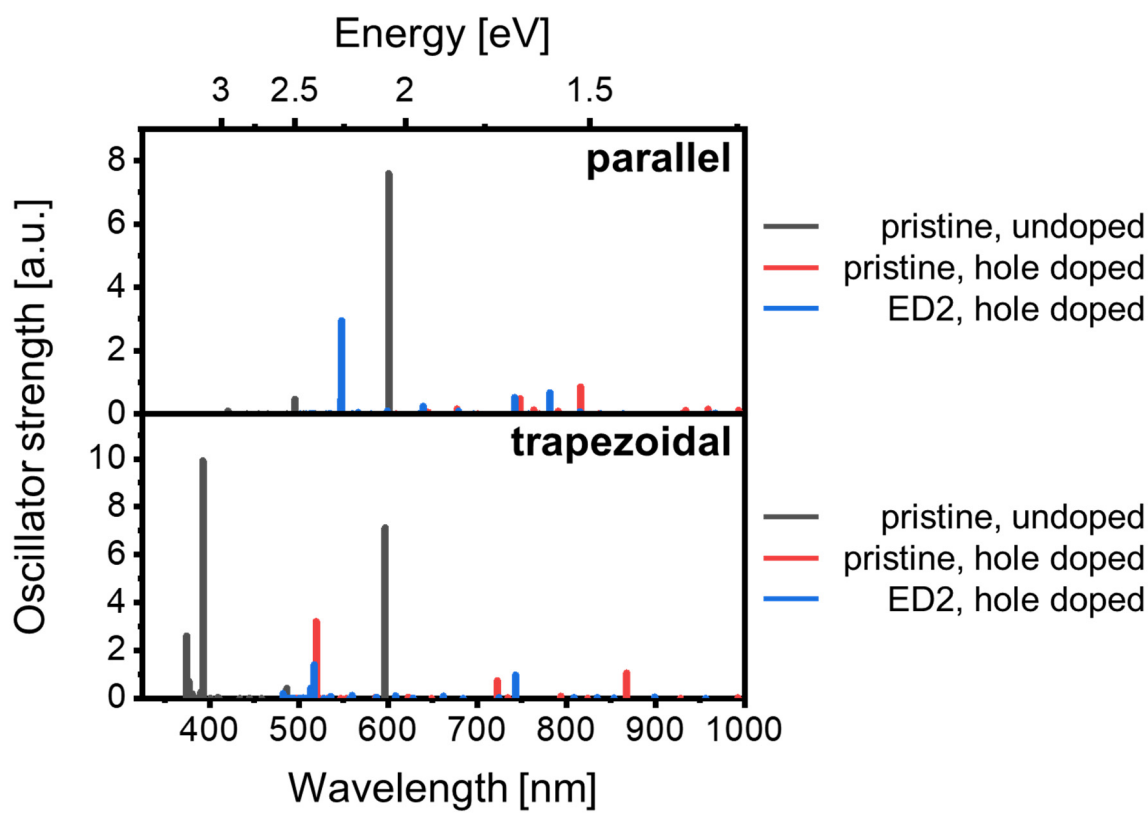

**Figure S21**: Comparison between the TD-UDFT (UωB97X-D/6-31G*) transition energies (unscaled data) computed for the parallel (top) and trapezoidal (bottom) 9-aGNRs for their pristine undoped (neutral, black line), pristine doped (charged, +1, red line) and defective ED2 doped species (charged, +1, blue line). The structural models considered refer to 16 terphenyl units for the parallel shape, and 15 units for the trapezoidal shape.